# Analog control of electrical conductivity in La$_{0.5}$Sr$_{0.5}$FeO$_{3-\delta}$ through oxygen deficiency induced magnetic transition

*Paul Nizet, Francesco Chiabrera\*, Nicolau López-Pintó, Nerea Alayo, Philipp Langner, Sergio Valencia, Arantxa Fraile-Rodríguez, Federico Baiutti, Alevtina Smekhova, Alex Morata, Jordi Sort, Albert Tarancón\**

Paul Nizet, Francesco Chiabrera, Nerea Alayo, Philipp Langner, Federico Baiutti, Alex Morata, Albert Tarancón
Department of Advanced Materials for Energy Applications, Catalonia Institute for Energy Research (IREC), Jardins de les Dones de Negre 1, 08930, Sant Adrià del Besòs, Barcelona, Spain
E-mail: fchiabrera@irec.cat , atarancon@irec.cat

Nicolau López-Pintó, Jordi Sort
Physics department, Universitat Autònoma de Barcelona, 08193 Cerdanyola del Vallès, Spain

Sergio Valencia, Alevtina Smekhova
Helmholtz-Zentrum Berlin für Materialien und Energie, Albert-Einstein-Str. 15, 12489 Berlin, Germany

Arantxa Fraile-Rodríguez
Universitat de Barcelona, Departament de Física de la Matèria Condensada, Barcelona, 08028, Spain

Arantxa Fraile-Rodríguez
Universitat de Barcelona, Institut de Nanociència i Nanotecnologia (IN2UB), 08028 Barcelona, Spain

Jordi Sort, Albert Tarancón
Catalan Institution for Research and Advanced Studies (ICREA), Passeig Lluís Companys 23, 08010, Barcelona, Spain

Switchability of materials properties by applying controlled stimuli such as voltage pulses is an emerging field of study with applicability in adaptive and programmable devices like neuromorphic transistors or non-emissive smart displays. One of the most exciting approaches to modulate materials performance is mobile ion/vacancy insertion for inducing changes in relevant electrical, optical, or magnetic properties, among others. Unveiling the interplay between changes in the concentration of mobile defects (like oxygen vacancies) and functional properties in relevant materials represents a step forward for underpinning the emerging oxide iontronics discipline. In this work, electrochemical oxide-ion pumping cells were fabricated for an analog control of the oxygen stoichiometry in thin films of mixed ionic-electronic conductor $La_{0.5}Sr_{0.5}FeO_{3-\delta}$. We demonstrate over more than 4 orders of magnitude electronic conductivity control within the same crystallographic phase through the precise and continuous voltage control of the oxygen stoichiometry. We show that behind the modification of the transport properties of the material lays a paramagnetic-to-antiferromagnetic transition. We exploit such magnetoelectric coupling to show control over the exchange interaction between $La_{0.5}Sr_{0.5}FeO_{3-\delta}$ and a ferromagnetic Co layer deposited on top.

## 1. Introduction

The tunability of specific material functional properties when exposed to different external conditions has drawn considerable attention in recent years. The external stimuli can be diverse, including temperature, pressure, incident light or electromagnetic fields.[1–3] More specifically, thin films of Transition Metal Oxides (TMOs) have shown to exhibit such a tunability with respect to heterogenous catalytic activity[4–6] as well as changes in the electronic structure,[7] in turn promoting the emergence of exciting phenomena such as high-temperature superconductivity, Colossal Magnetoresistance (CMR), or Metal-to-Insulator Transitions (MIT).[8–10] Modulation capabilities make thin film of TMOs potential materials for novel and advanced switchable devices such as neuromorphic transistors, memristors, and Resistive Random-Access Memories (Re-RAMs) that strongly depend on a precise tuning of the functional properties.[11–14]

An excellent example of the modulation of functional properties in TMOs can be found in the perovskite oxide $La_{1-x}Sr_xFeO_{3-\delta}$ (LSFx, x=0-1) family, a material system that finds multiple applications e.g. in Solid Oxide Fuel Cells (SOFC) electrodes,[15] catalyst for Oxygen Evolution Reaction (OER) [6,16] or as functional materials in advanced computing devices.[17] In the LSFx compounds, control of electronic, optical and catalytic properties can be achieved by two strategies: i) aliovalent $Sr^{2+}$ substitution of trivalent $La^{3+}$ ($Sr'_{La}$ in Kroner-Vink notation), and ii) variation of oxygen deficiency δ. On the one hand, Sr substitution gives rise to a progressive increase of electron hole concentration, enabling a wide range of property change from an insulating and semi-transparent material for x=0 ($LaFeO_3$, LFO) to metallic and opaque for x=1 ($SrFeO_3$, SFO).[18] On the other hand, the introduction of oxygen deficiency tends to reestablish the insulating state, since the introduction of positively charged oxygen vacancies ($V_O^{\bullet\bullet}$) is naturally compensated by a reduction of electronic holes.[19] Engineering the functionalities of LSFx relies on controlling the acceptor dopant or the oxygen vacancy content. Notably, while the former is fixed during synthesis, the latter allows tunability during operation, potentially offering an analog control of the majority charge carrier concentration and the electronic properties.

However, depending on the $Sr^{2+}$ substitution, different behaviors have been observed when reducing LSFx. For x>0.7, LSFx experiences a topotactic phase transition between the oxidized perovskite (PV) and the reduced Brownmillerite (BM) crystal structure, characterized by the formation of long-range oxygen vacancy ordering.[20,21] Electronic, optical and thermal conductivity varies substantially across this transition, with the BM phase of SFO (i.e. $SrFeO_{2.5}$) presenting an insulating behavior similar to undoped LFO.[7,20–22] For x≤0.7, the BM transition is not usually reported, in contrast to other similar materials such as $La_{1-x}Sr_xCoO_{3-\delta}$ (LSC),[23] LSFx is expected to crystallize in perovskite structure with a random distribution of oxygen vacancies.[20,21] In this range, electronic transport properties as a function of the oxygen deficiency are not fully understood. Several studies carried out at high temperatures (T >500ºC) concluded that electronic conductivity varies almost linearly with electronic hole concentration (i.e. majority charge carriers).[24–26] This model, however, fails to predict the reported exponential variation of electronic conductivity with hole concentration in the intermediate-to-low temperature regime (T <500ºC).[18,19,27,28] The origin of the discrepancy between high and intermediate-to-low temperature conductivity measurements is still under debate. Some hypotheses suggest a blocking effect of the Fe-O-Fe conduction pathway induced by vacancy association mechanisms.[29] Density Functional Theory (DFT) calculations showed that oxygen vacancy might increase electronic hole localization [30] and its effective mass,[9] affecting the electronic transport. Understanding the effect of oxygen stoichiometry in PV LSFx may offer an advanced analog control of its functional properties (i.e., fine-tuning of properties within a wide, rather than discrete or binary, range of values) since, contrary to the BM/PV transition, oxygen ion concentration can be varied in a precise and continuous fashion.

Currently, the two main strategies to study functional properties as function of oxygen stoichiometry are $pO_2$ equilibration during deposition and annealing under controlled atmosphere.[7,23,31,32] Despite these techniques succeed in efficiently changing the oxidation state of thin film materials, they do not offer the precise control of the oxygen concentration. Moreover, temperature dependence studies are not possible since the oxygen equilibrium with the atmosphere, i.e., the final oxygen content in the material, shifts with temperature.[33] Recent studies have shown the possibility of using voltage gating through an electrolyte to modulate, in a reversible manner, the oxygen stoichiometry by fine tuning the oxygen chemical potential. Following this procedure, several studies have in-situ characterized the conductivity, magnetic and thermal properties of different oxides.[22,34,35]

In this work, we present a comprehensive study of the conductivity changes in $La_{0.5}Sr_{0.5}FeO_{3-\delta}$ (LSF50) as a function of oxygen deficiency and temperature (T= 50–350°C). A voltage-modulation approach is followed to enable reversible switching of oxygen stoichiometry in the LSF50 thin films while carrying out an in-situ characterization of the electronic functional properties. Sr-doping concentrations of x= 0.5 is studied to provide a broad understanding of the role of the oxygen defects preserving the PV structure, i.e. excluding the effect of crystallographic phase transitions. To elucidate the origin of the changes in conductivity with oxygen vacancy concentration all along the range of conditions, electrical measurements are combined with optical (Spectroscopic Ellipsometry) and magnetic (Vibrating-Sample Magnetometry and Spin-resolved Photoemission-Electron Microscopy) characterization. According to a direct interrelation obtained between functional properties and defect concentration, the modulation of the electronic properties of LSF can be understood as a consequence of spin-order effects in these materials. We show that the voltage control of the oxygen stoichiometry in LSF50 can be used to manipulate the magnetic state of a cobalt ferromagnetic layer deposited on top, thus demonstrating the potential interest of LSF50 as an active material in magnetoionic field.

## 2. Results and Discussion

### 2.1. LSF50 conductivity modulation under applied voltage

The controlled modulation of electrical conductivity in LSF50 thin films was pursued by electrochemically pumping oxygen ions through an oxide-ion electrolyte. For this reason, an LSF50 thin film was deposited by Pulsed Laser Deposition (PLD) on 20 nm gadolinium-doped ceria (CGO)-coated Yttria Stabilized Zirconia (YSZ) (001) oxide ion-electrolytes (see Experimental Section for more details). Structural and surface characterization revealed smooth layers of 100 nm in thickness and preferential out of plane (00l) orientation with a 45º in-plane rotation compared with the substrate fluorite unit cell (see **Supporting Information Sections S1 and S2**). **Figure 1a** shows a sketch of the configuration used to measure the electrical conductivity of a LSF50 layer during voltage-controlled oxygen ion insertion/extraction. The gating voltage (Vg) is applied across the YSZ substrate acting as an electrolyte. Four conductive stripes of 10 nm Ti/100 nm Au were microfabricated by lift-off and metal thermal evaporation to act as current collector and to perform 4-point contact conductivity measurements. Additionally, a capping layer of alumina ($Al_2O_3$) of approximately 100 nm was also deposited via PLD on top of the LSF50 thin film to prevent the equilibrium of the functional layer with the atmosphere. On the backside of the YSZ electrolyte, silver paste was painted to act as counter electrode (CE). The samples were then heated up till 350 ºC and an electrochemical potential was applied to the film with the CE grounded ($V_G$=-0.45 V to 0.1V). To ensure a proper stabilization of the material, the current flowing through the system ($I_G$) was measured (**Figure 1b** and **Supporting Information Section S3**). Due to the pure oxide-ion conducting properties of YSZ substrates, this current represents the oxygen ions flowing from the CE, where oxygen incorporation/evolution is promoted by the silver paste, to the LSF50 layer. As seen in **Figure 1b**, after each voltage step, $I_G$ current goes to zero, indicating that the oxygen ion flow is correctly blocked by the alumina capping layer and electrochemical equilibrium at that specific potential is achieved.

The electrical conductivity (σ) of the LSF50 for different equilibrium states was then characterized in a two-steps process consisting of: (1) state setting by application of an electrochemical potential at 350 ºC ($V_G$=-0.45 V to 0.1V) and (2) temperature-dependent 4-point contact conductivity measurements (T=50-350 ºC) under open circuit voltage (OCV) conditions ($I_G$=0, see inset of **Figure 1c**). The two-steps process was repeated for all the different gate voltages ($V_G$=-0.45 V to 0.1V). Importantly, the non-volatility of the electrochemical state and the reproducibility of the measurements was assessed by measuring a constant conductivity and OCV after repeated cycles of $V_G$, temperature, and time (see **Supporting Information Section 3**). **Figure 1c** shows the equilibrium conductivity presented as a function of $V_G$ applied at different temperatures. Precise modulation of the conductivity with the voltage was achieved over more than four orders of magnitude in the whole range of temperatures, e.g. σ=3·$10^{-3}$ S/cm at -0.4 V to σ=1·$10^2$ S/cm at 0V when cooling down to T=50ºC after the state setting. As can be seen, the material presents a continuous increase in conductivity starting at negative potentials until reaching a plateau when approaching $V_G$= 0V. The next sections are dedicated to elucidating the effect of ion insertion into the LSF50 material.

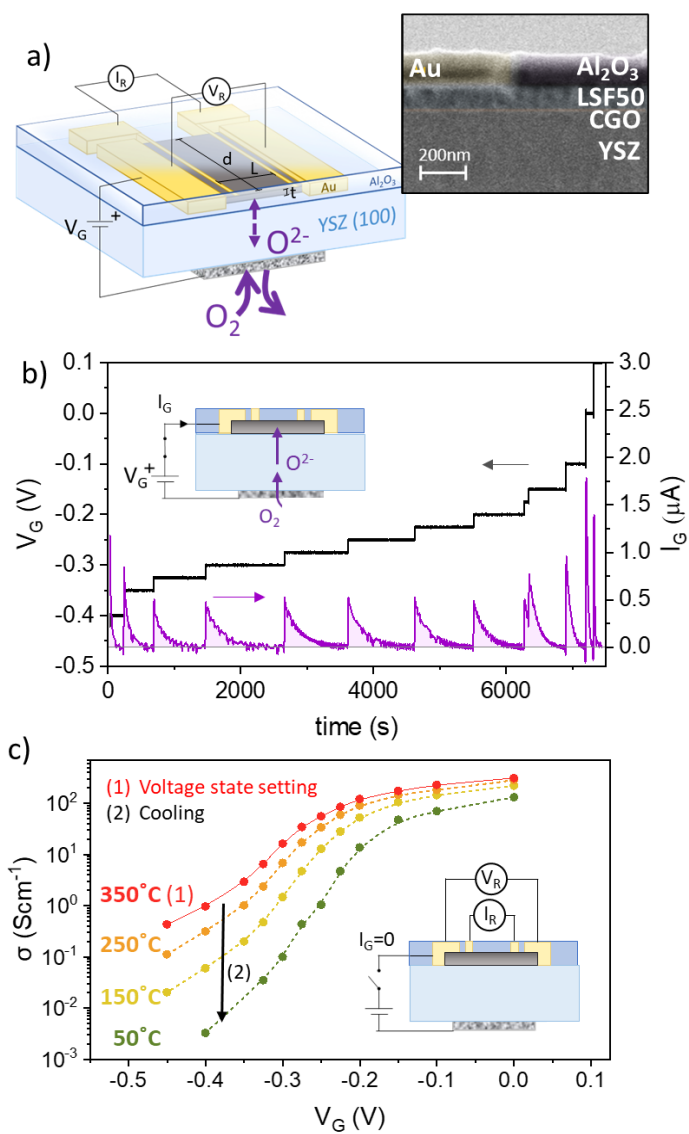

**Figure 1.** a) Sketch of the sample showing the main dimensions: thickness (t), width (d) and distance between pads (L); and the four-probe configuration for in-plane measurement of the electrical resistance and out-of-plane electrochemical pumping cell (Au/YSZ/Ag) for oxygen ion insertion in the LSF50 film. Purple arrows indicate the (de)insertion path of O into the film. The inset illustrates a cross-section SEM view of the sample including the alumina capping. b) Current stabilization of the LSF50 sample after the application of different gate voltages $V_G$. c) Conductivity measurements of the LSF50 layer after application of different $V_G$ applied at 350ºC and subsequent cooling to different temperatures.

## 2.2. Control of the defect concentration in LSF50 with the applied voltage

First, the quantification of the oxide-ion insertion into LSF50 as a function of the applied voltage was carried out. In this regard, the application of an electrochemical potential ($\Delta E$) to an oxide thin film through an oxide-ion electrolyte gives rise to a variation of its oxygen chemical potential, analogous to a modification of the equilibrium oxygen partial pressure ($pO_{2,eq}$) through the Nerst potential.[36,37] This shift modifies the defect equilibrium in the thin film leading to an increase/decrease of its oxygen concentration, similarly to batteries.[38] In the case of LSF50 thin films, the equilibrium with oxygen is governed by the following defect chemistry equation:[36,37]

$$\frac{1}{2}O_2 + V_O^{\bullet\bullet} + 2Fe_{Fe}^x \leftrightarrow O_O^x + 2Fe_{Fe}^{\bullet} \qquad (1)$$

where $Fe_{Fe}^x$, $O_O^x$, $V_O^{\bullet\bullet}$, and $Fe_{Fe}^{\bullet}$ denote the $Fe^{3+}$, oxygen ions, oxygen vacancies and $Fe^{4+}$ holes species respectively, according to the Kroger-Vink notation. In this equation, the incorporated oxygen atoms occupy vacant oxygen sites and the iron species, present in the structure initially in the form of $Fe^{3+}$, undergo oxidation to $Fe^{4+}$. Iron oxidation states, such as $Fe^{2+}$ and $Fe^{5+}$, can be neglected within the range of temperature and oxygen partial pressure of this study.[37,38] Indeed, it is worth mentioning that the generalized treatment of the electronic holes as $Fe_{Fe}^{\bullet}$ is already a simplification as holes localize in the form of polarons over the Fe-O bond.[18,27,30] Two electron holes are formed upon incorporation of a single oxygen atom, i.e. $[Sr'_{La}]=2\delta+[Fe_{Fe}^{\bullet}]$. As a consequence, the change of oxygen stoichiometry induced when modifying oxygen equilibrium through $V_G$ involves a current ($I_G$) flowing through the system (see **Figure 1b**). The integration of this current along the transition between stable states can be used to calculate the change in oxygen stoichiometry ($\Delta\delta$) in the material (see **Figure 2a**):[39]

$$\Delta\delta = -\frac{c_{LSF}^3}{2eV_{film}} \int_0^t I_G(t)dt \qquad (2)$$

where $c_{LSF}$ is the lattice parameter and $V_{film}$ is the volume of the film. Considering that according to previous studies the samples were entirely reduced under the lowest $pO_{2,eq}$ (i.e. $[Sr'_{La}]=2\delta$ and $[Fe_{Fe}^{\bullet}] = 0$),[36,37] one can calculate the state of oxidation of the iron in the material and the oxygen content as a function of the applied voltage. This assumption and the obtained stoichiometry values were confirmed by employing in-situ ellipsometry using a methodology recently developed by the authors (see **Section S4 in Supporting Information**).[36]

**Figure 2b** depicts the oxidation state of iron (and oxygen content 3- δ) in the LSF50 sample as a function of the $pO_{2,eq}$ (and $V_G$) at 350ºC. The material reaches a maximum non-stoichiometry of δ=0.25 and follows a sigmoidal-type shape behavior expected from the dilute model proposed by Mizusaki et al.[24] and described by Equation 1 (assuming non-interacting oxygen vacancies, see **Section S3 in Supporting Information**). In the case of the experimental data, the fit of the oxidation curve with this model yields to an equilibrium oxidation constant of Equation 1 of $K_{ox}$=5000, which falls within the expected range.[24,36,37] The slight deviations from the fit when approaching the maximum oxidation state are likely due to non-dilute interaction, as reported by Tang et al.[36] This good agreement with the defect chemistry model indicates that the oxygen content in LSF50 thin films follows the expected behavior of a single-phase solid solution system. This is also confirmed by ex-situ X-Ray diffraction (XRD) analysis, which revealed that the structure expanded upon reduction but retaining its PV phase (see **Supporting Information Section S2**). Notably, **Figure 2c** shows that the increase of lattice parameter is consistent with the variation of Fe-O bond length predicted by a simple hard sphere approximation as the smaller $Fe^{4+}$ (58.5 pm) is substituted by the larger $Fe^{3+}$ ions (64.5 pm).[40,41] The consistency with the lattice parameters measured in literature for fully oxidized LSFx with different Sr content suggests that the iron oxidation state presents the main influence on the lattice parameter of the PV phase.[42,43]

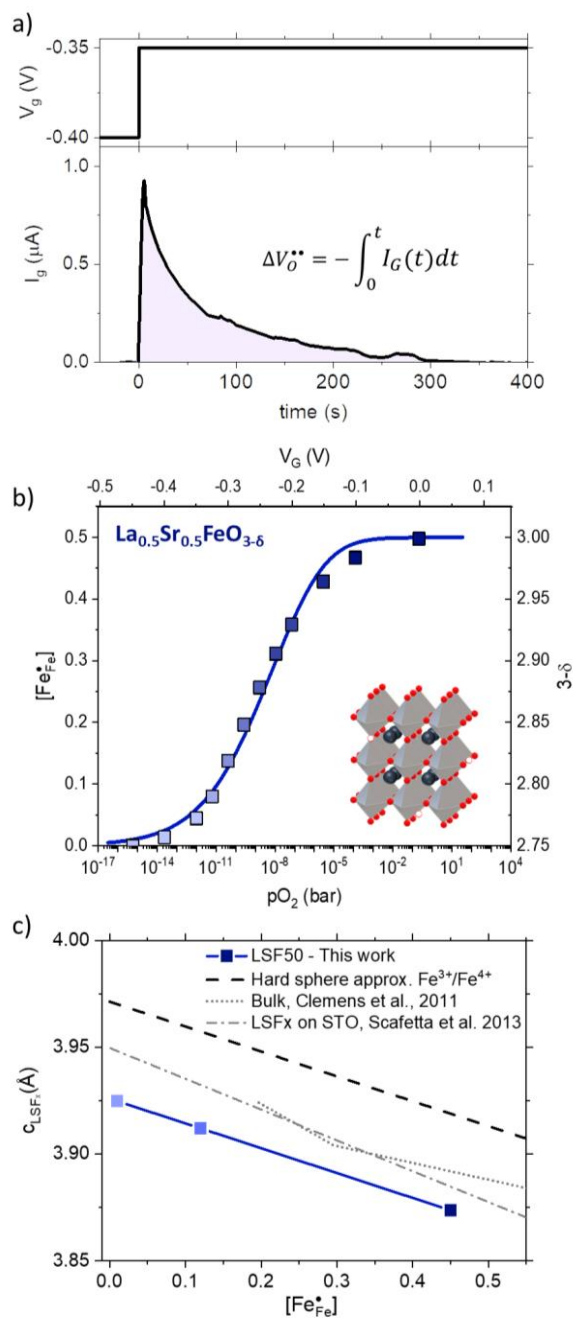

**Figure 2.** a) Exponential decay of the current after application of a voltage step. Its integration allows for the calculation of the charge and, consequently, the final oxidation state. b) Oxidation curve extracted from current integration for LSF50. The figure also includes the relation between $[O_O^x]$=3-δ and $[Fe_{Fe}^\bullet]$. The inset is a schematic of the PV structure, symbolizing the single-phase solid solution model employed for the fitting shown in blue line. c) Calculated cell parameter from ex-situ XRD of the LSF50 samples after the application of different $V_G$ at 350ºC indicated by the calculated $[Fe_{Fe}^\bullet]$. Reference measurements on LSFx cell parameters are also shown for comparison in grey dotted and dash-dotted lines.[42,43]

## 2.3. The effect of oxygen deficiency on the electrical and optical conductivity

To understand the origin of the changes in the electronic transport of LSF50, the conductivity of the layers was measured as a function of the concentration of charge carriers ($Fe_{Fe}^{\bullet}$) at constant temperature in the range of T=50-350°C (see **Figure 3**). Importantly, this analysis is possible because of the battery-like configuration of our system and the Al$_2$O$_3$ capping layer, which ensures that under OCV (I$_G$=0) the sample does not equilibrate with the atmosphere and retains its defect concentration upon time and temperature cycling (see **Supporting Information Section 3**). The evolution of electrical conductivity shown in **Figure 3a** appears to change exponentially with the hole concentration [$Fe_{Fe}^{\bullet}$] and cannot be simply explained by the variation of the charge carriers. Moreover, decreasing the temperature is observed to affect the conductivity of oxygen deficient LSF50 more than the oxidized one, suggesting an intricate behavior between charge carrier and mobility.

To investigate this point, conductivity results are displayed in an Arrhenius-like representation for different [$Fe_{Fe}^{\bullet}$], see **Figure 3b**. The conductivity exhibits a linear relationship in ln(σT) vs 1/T for the most reduced states and oxidized states suggesting that the electronic transport in the thin film can be characterized by adiabatic small polaron hopping conduction. Electronic conductivity of LSF is commonly described by the polaron hopping model where the holes, localized over the Fe-O bond, can move to the Nearest Neighbor Hopping site with the assistance of phonons, [18,28,33] i.e. showing a thermal activation that follows an Arrhenius law with a certain activation energy ($E_a$) and pre-exponential factor ($\sigma_o$):[44,45]

$$\sigma(T) = \frac{\sigma_o}{T} exp\left(\frac{E_a}{k_b T}\right), \quad \sigma_o = \frac{e^2 \nu}{a k_b}[Fe_{Fe}^{\bullet}] \cdot [Fe_{Fe}^{x}] \quad (3,4)$$

where e, $\nu$, $a$ are the electron charge, the polaron attempt frequency and the site-to-site hopping distance, respectively. The preexponential factor in this model is the isothermal conductivity term of the equation and includes the probability of finding an $Fe_{Fe}^{\bullet}$ next to an $Fe_{Fe}^{x}$ site. Although the conductivity can be described with the polaron model for the most reduced and oxidized states, there is a change in the slope between the reduced and oxidized states of LSF50, strongly suggesting a close relationship between the activation energy and the hole concentration. The trend was equally followed with the non-adiabatic polaron hopping model leading to the same analysis outputs (see **Supporting Information Section 5**). To analyze this change, **Figure 3c and 3d** illustrate the variation in $E_a$ and $\sigma_o$ as a function of the hole concentration fitted over different temperature ranges. The activation energy starts from a constant value of 0.35 eV for the fully reduced material and then progressively decreases to a value of 0.1 eV as the material oxidizes. This decrease does not happen simultaneously for all the temperatures, but there is a strong temperature dependance on the precise $[Fe_{Fe}^{\bullet}]$ values at which this $E_a$ decrease occurs. Actually, when examining a specific oxidation state, the $E_a$ diminishes as the temperature increases. A clear illustration of this transition can be seen at $[Fe_{Fe}^{\bullet}]=0.25$, where the activation energy is 0.35 eV at low temperatures and gradually decreases, reaching 0.15 eV at 325 °C. A non-trivial behavior in the $\sigma_o$ evolution in **Figure 3d** is also seen. At the highest measured temperatures, small variations in the preexponential factors are seen. Nevertheless, low temperature $\sigma_o$ plots show an exponential increase with $[Fe_{Fe}^{\bullet}]$. This exponential evolution of $\sigma_o$ is followed up to a given $[Fe_{Fe}^{\bullet}]$ in which it suddenly decreases back to values close to $10^6$ S·K·cm$^{-1}$. The exact $[Fe_{Fe}^{\bullet}]$ at which the maximum of $\sigma_o$ is observed matches well with the onset of the decrease of $E_a$. This suggests a relation between the high activation energy states ($E_a= 0.35$ eV) and the exponential increase of $\sigma_o$ with $[Fe_{Fe}^{\bullet}]$ ($\sigma_o \alpha\ e^{[Fe_{Fe}^{\bullet}]}$).

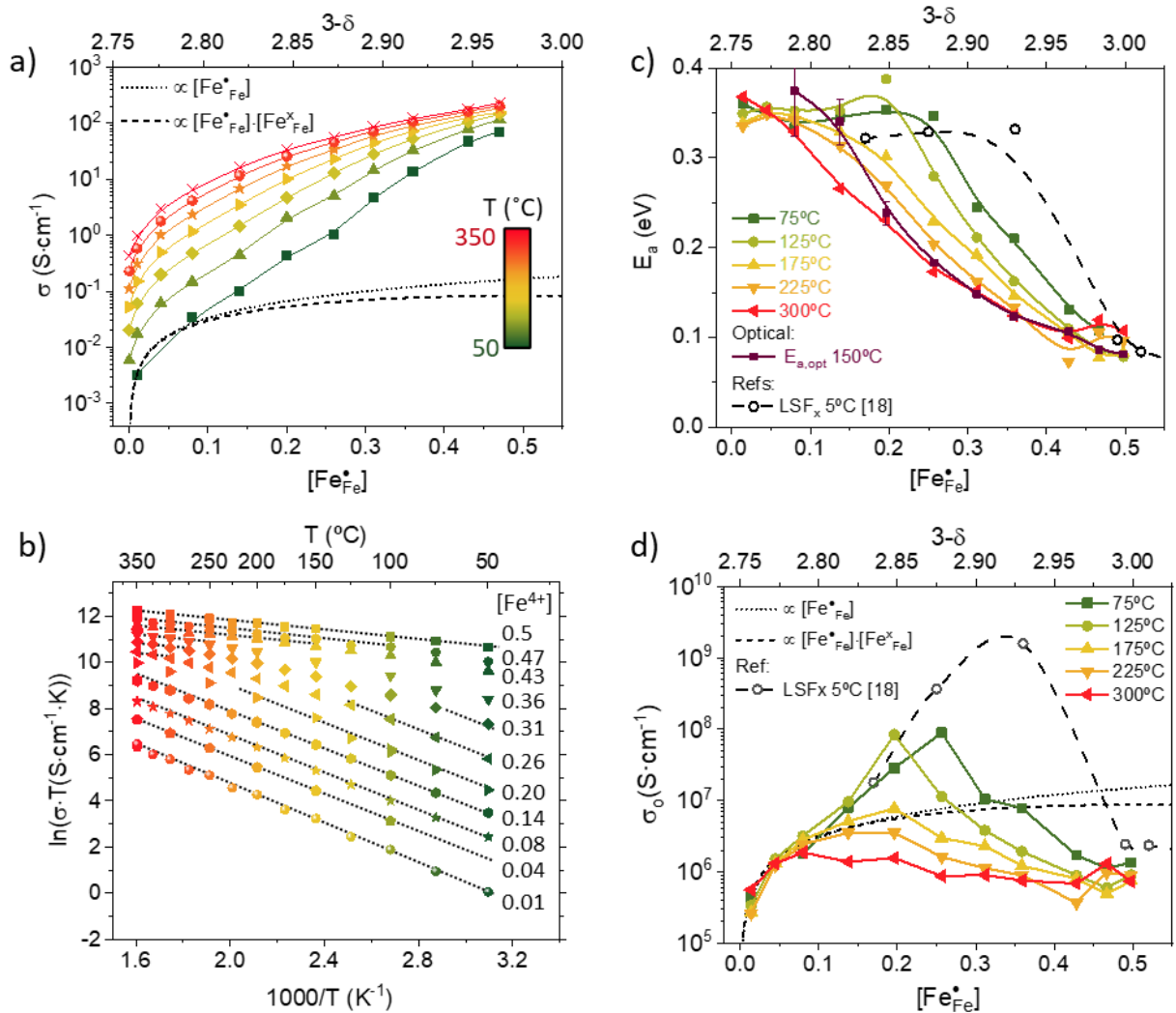

**Figure 3.** a) Conductivity of LSF50 as a function of the hole concentration for different temperatures. Solid line shows the expected change in conductivity according to the polaron hopping model. b) Arrhenius plot of the LSF50 conductivity at different oxidation states. The numbers on the right indicate the $[Fe_{Fe}^{\bullet}]$. c-d) Activation energy ($E_a$) and preexponential factor ($\sigma_o$) dependance on the temperature and oxidation state extracted from the Arrhenius fitting of the conductivity. The dashed line corresponds to Xie et al. study.[18] $E_{a,opt}$ extracted from the optical measurements at 150C is also plotted.

Up until now, the results suggest a thermally activated transition, which remarkably lowers the activation energy for polaron hopping. The origin of this transition cannot be attributed to a phase change towards BM, as no structural transition was observed in the films (see **Supporting Information Section 2**). Phase mixing models also failed to fit the LSF50 data (see **Supporting Information Section 8**) as no mixture of a highly conductive and insulating regions is expected in this single-phase material.[46] Interestingly, a similar decrease of activation energy in the LSFx family had been documented in studies involving fully oxidized samples with varying Sr contents ranging from 0.17 to 1.[18,47] As a matter of comparison, the values extracted from the study of Xie et al.[18] as a function of Sr concentration (assuming a fully oxidized $[Sr_{Sr}'] = [Fe_{Fe}^{\bullet}]$ equilibrium) for a lower T range (5 °C) are also plotted in **Figure 3c**. The evolution of the preexponential factor is also shown **Figure 3d**. The trend of both sets of data match for the most reduced and oxidized region, both for $E_a$ and $\sigma_o$. However, the $[Fe_{Fe}^{\bullet}]$ at which the $E_a$ starts to decrease and $\sigma_o$ presents its maximum is higher in the study of Xie et al. compared to our data, which delineates a trend with the T-range investigated (see values extracted at lower T in **Supporting Information Section 6**). The comparison suggests that the conductivity is mainly dependent on the hole concentration $[Fe_{Fe}^{\bullet}]$ and on the T range, regardless if modified by the aliovalent doping concentration or the oxygen deficiency, therefore excluding strong interaction between holes and other point defects ($V_O^{\bullet\bullet}$ or $Sr_{La}'$) or other mechanisms of oxygen vacancies ordering.

Previous studies have correlated the hole concentration in LSFx to modifications of its band structure, which could explain the transition from a conductive to insulating state.[18,27,30] To find a direct correlation with the electrical transition, in-situ spectroscopic ellipsometry measurements were carried out as a function of the oxidation state and temperature (see **Experimental** and **Supporting Information Section 4**). **Figure 4a** displays the optical conductivity for different $[Fe_{Fe}^{\bullet}]$ at a fixed temperature of 350°C. Starting with the most reduced state and progressing towards the more oxidized ones, an increase in the intensity and a red shift of the infrared (A) and visible (B) transitions at the expense of UV transition (C) is observed (see **Supporting Information SI7** for a more detailed description on the band structure evolution of the LSFx family of materials). This progressive evolution of the band structure is visible in the decrease of the onset energy of transitions A and B upon oxidation (**Figure 4b**), calculated by direct-forbidden Tauc-plots (see **Figure S8** in **Supporting Information**). The trend observed for transition B well matches with the literature data of oxidized LSFx thin films with different Sr content when plotted as a function of $[Fe_{Fe}^\bullet]$, confirming that electronic holes are the main responsible for band structure modification.[27] Although this band-closing goes in line with the increase of electronic conduction, we note that the optical bandgap of B is not affected by temperature, in disagreement with what observed for the onset of the electronic transition, see **Figure 3**. On the contrary, the bandgap of A presents a clear variation with temperature.

To gain further insights into this behavior, **Figure 4c** reports the deconvoluted optical conductivity of transition A measured for three different $[Fe_{Fe}^\bullet]$ as a function of T (see **Supporting Information SI4** for details on the deconvolution). In the most oxidized case, $[Fe_{Fe}^\bullet]$= 0.5, the shape and position of the peak barely change, while for intermediate states ($[Fe_{Fe}^\bullet]$= 0.26 and $[Fe_{Fe}^\bullet]$= 0.2) a progressive red shift of the peak energy is observed as the T is increased, in agreement with the behavior measured for $E_a$. As the results suggest a relation between the polaron hopping energy and transition A, we tested this hypothesis in the frame of the optical adiabatic small polaron model developed by Holstein.[45] According to this theory, transition A is related to the photon energy ($\hbar\omega_p$) required to induce polaron excitation from one Fe-site to its adjacent counterpart, following the equation $\hbar\omega_p = 2E_p$, where $E_p$ is the polaron formation energy.[48,49] Considering the adiabatic approximation, the relation between polaron hopping energy and $E_p$ becomes:[45]

$$E_{a,opt} = \frac{1}{2}E_p - J = \frac{1}{4}\hbar\omega_p - J \qquad (5)$$

wherein $J$ represents the transfer energy, typically hovering around 0.1 eV.[49] Together with the experimental $E_a$ obtained for electrical measurements, **Figure 3c** includes the activation energy extracted from the optical measurements considering $J$= 0.1 eV (see more details in **Supporting Information Section 4**). The evolution of $E_{a,opt}$ closely follows the electrical $E_a$, revealing the polaron hopping nature of transition A. The good agreement between optical and electrical measurements shows that the evolution of polaronic characteristics can be tracked by optical conductivity.

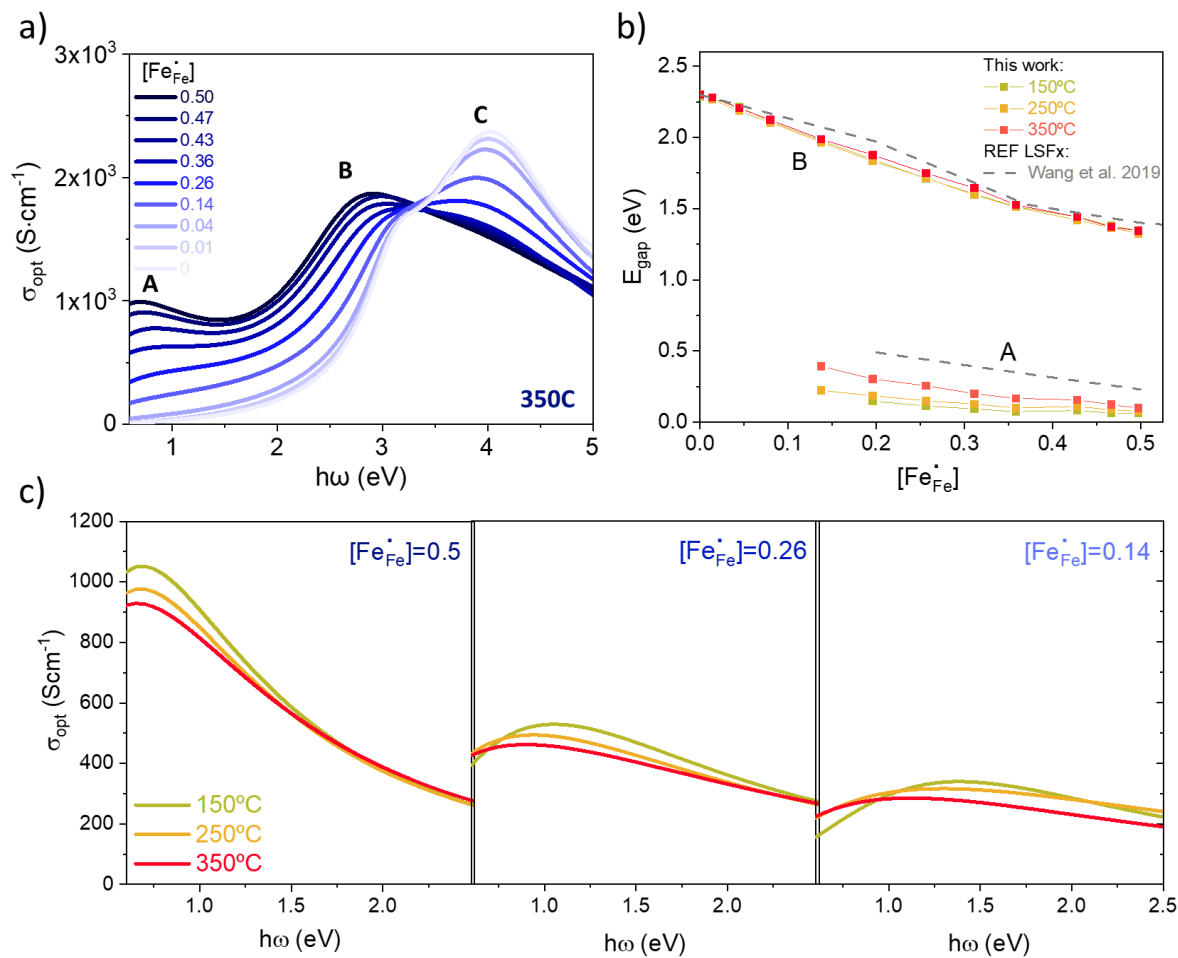

**Figure 4.** a) Calculated optical conductivity for the different measured oxidation states indicated by their $[Fe^{4+}]$ at 350C for LSF50. b) Calculated band gap closing as a function of $[Fe^{4+}]$. The figure also includes the band evolution of different fully oxidized LSFx from Wang et al.[27] c) Optical conductivity of the first transition peak at three different oxidation states indicated by the $[Fe^{4+}]$ measured at different temperatures.

Overall, the optical analysis reveals that the evolution of the band-structure upon oxygen variation, represented by the onset of B, cannot account for the thermal behavior observed in the electrical behavior. Transition A is instead able to properly reproduce the electrical transition, which can be explained in the frame of the optical adiabatic small polaron model. Therefore, combined optical and electrical measurements show that, for a constant electronic hole concentration, the activation energy tends to decrease while increasing the temperature, highlighting the thermal sensitivity of the transition. In this regard, electron transport in oxides, along with other properties, is very sensitive to the type of interaction between B-site cations through oxygen atoms.[9,50–52] Given the spin-order relation to these exchange interactions, the magnetic properties of LSF50 were also investigated.

## 2.5. The impact of magnetic properties on the electronic conductivity of LSF50

Considering the parent compound LaFeO$_3$, a super-exchange (SE) interaction between neighboring Fe$^{3+}$ results in the establishment of an antiferromagnetic (AFM) interaction, characterized by a Néel Temperature ($T_N$) of 430 ºC.[53] Mixed valence Fe$^{3+}$/Fe$^{4+}$ promotes instead the ferromagnetic (FM) spin alignment through double-exchange (DE) interactions, resulting in FM or paramagnetic state (PM) for aliovalent Sr substitution.[51–54] Within this scenario fully oxidized LSF50 is paramagnetic. The reduction of the oxygen content is expected to promote and AFM phase.

The magnetic domain state of two 100nm-LSF50 layers, one previously reduced and the other previously oxidized electrochemically at 350ºC and quenched to room temperature, has been imaged by means of X-ray photoemission-electron microscopy (PEEM) using X-ray magnetic linear dichroism (XMLD) as magnetic contrast mechanism. XMLD is proportional to the square of the magnetic moment and hence commonly used for the investigation of AFM compounds.[55–57] Since no alumina capping could be used due to the low probing depth of XAS, the oxidation state of the samples was measured ex-situ by ellipsometry ($[Fe^{\bullet}_{Fe}]$= 0.04 and 0.5 for the reduced and oxidized, respectively). **Figure 5a and 5b** show the XMLD images obtained at the Fe $L_2$ edge, see **Experimental section**, for the reduced and oxidized samples, respectively. A clear pattern arising from the AFM domains is observed for the reduced sample. The AFM domain size in the reduced LSF50 is in good agreement with the previously documented in LaFeO$_3$ corroborating the role of SE interaction in AFM order in the LSF family of materials.[55–57] On the contrary, no domains were detected within the oxidized region irrespective of the angle of the incoming linearly polarized radiation with respect to the samples surface, in agreement with a paramagnetic phase.

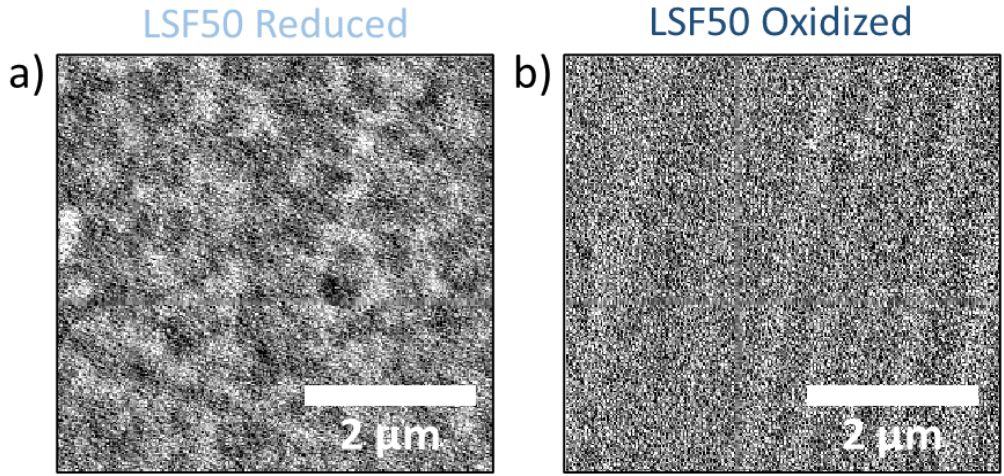

**Figure 5** XMLD-PEEM images of antiferromagnetic domain patterns recorded for the reduced **(a)** and oxidized **(b)** LSF50 samples.

Having assessed the antiferromagnetic/paramagnetic (AFM/PM) transition in LSF50 thin films upon oxidation, we further explored the relation between magnetic and electrical properties. First, the $T_N$ was obtained from the temperature dependence of the magnetization (M) for different oxidation states ($[Fe^{\bullet}_{Fe}]$ =0.1, 0.2, 0.3, 0.46). The measurements were performed at a constant field (H=1kOe) as shown in **Figure S16**. A change in the M curve was observed in all oxidation states that was assigned to the AFM/PM transition. The $T_N$ of the sample was determined as a function of $[Fe^{\bullet}_{Fe}]$ (see **Figure S16**). **Figure 6a** shows the $T_N$ values extracted from the measurements. Although the precision is not high due to the wide temperature span of the transition, the results show good agreement with bulk measurement of $T_N$ carried out by Wattiaux et al.[53] as a function of $[Fe^{\bullet}_{Fe}]$, also presented in **Figure 6a**. The temperature-span of the electrical transition was also derived from **Figure 3c**, considering the temperatures at which the $E_a$ sets at its resistive state ($T_{Ea,set}$) and at its conductive state ($T_{Ea,On}$) (see **Supporting Information Section 6** for the procedure). Their values are included in **Figure 6a** as a function of the $[Fe^{\bullet}_{Fe}]$. Both $T_{Ea,set}$ and $T_{Ea,On}$ show a direct relation with hole concentration, as they decrease linearly as $[Fe^{\bullet}_{Fe}]$ rises. Moreover, a clear match in both trend and absolute value between $T_N$ and $T_{Ea,On}$ is evident, which also includes the $T_{Ea,On}$ extracted from Xie et al. [18] and previously reported in **Figure 3c**. It is important to highlight that the $T_{Ea,On}$ and $T_{Ea,set}$ points from Xie et al. [18] and the $T_N$ as a function of $[Fe^{\bullet}_{Fe}]$ relationship by Wattiaux et al.[53] were acquired using LSF samples with different Sr concentrations. In contrast, the measurements presented in this work have been performed using the same material while modifying $[Fe^{\bullet}_{Fe}]$ through adjustments of the oxygen vacancy concentration. This emphasizes once again that the change of $[Fe^{\bullet}_{Fe}]$ in LSF50 is the true responsible for the change in the material properties.

The observed correspondence between $T_{Ea}$ and $T_N$ suggests a correlation between the magnetic order of the material and its activation energy. As the polaron hopping takes place through a Fe-O-Fe bond, the two main interactions play an important role in both electrical and magnetic response. Indeed, the elevated $E_a$ associated with low $[Fe^{\bullet}_{Fe}]$ arises from spin restrictions inherent to SE interactions a mechanism also responsible for the AFM ordering. Hence, owing to spin constraints, hopping preferentially occurs within the nearest neighbor with the same spin orientation, leading to a hop to the second nearest neighbor (SNN) Fe site rather than the first (see AFM paths in **Figure 6b**). Indeed, DFT calculations on SNN hopping in LSF coincide with the experimental $E_a$=0.35eV measured in this work.[58] As expected for systems with competing DE and SE interactions, the increase of $[Fe^{\bullet}_{Fe}]$ leads to a weakening of the effective SE interaction, therefore lowering $T_N$.[51,52] When temperature reaches $T_N$, the thermal energy overcomes the preferential spin order and the hopping to the first nearest neighbor is favored (a process with a lower hopping energy, see PM in **Figure 6b**). This hypothesis also agrees with the shift in the A-transition energy, see **Figure 6b**. Therefore, the alignment of $T_N$ and $T_{Ea}$ is not coincidental but rather the main reason for the $E_a$ change observed. The magnetic transition also provides an explanation for conventional measurements carried out at elevated temperatures, where conductivity could be readily explained by the polaron hopping model, with an $E_a$ around 0.1 eV.[54] Since these measurements were performed at temperatures exceeding 500 ºC, LSF50 was always in its PM state, thus precluding the observation of spin ordering effects. Moreover, the exponential increase of $\sigma_o$ with $[Fe^{\bullet}_{Fe}]$ is then restricted and particular of the AFM state. We also assign the origin of this exponential increase in conductivity with $[Fe^{\bullet}_{Fe}]$ to the SE-DE competition. It is well known that the increase of $[Fe^{\bullet}_{Fe}]$ increases the Fe 3d/O2p orbitals hybridization.[27,30] The increase in the delocalization of the polaron could lead to a higher hopping probability.[48] This explains why $E_a$ maintains its value, as AFM order is preserved, but $\sigma_o$ is exponentially affected by $[Fe^{\bullet}_{Fe}]$. Summarizing, our results demonstrate that the magnetic properties of LSF50 ruled by the concentration of $Fe^{\bullet}_{Fe}$ is at the origin of the modulation of its electrical properties.

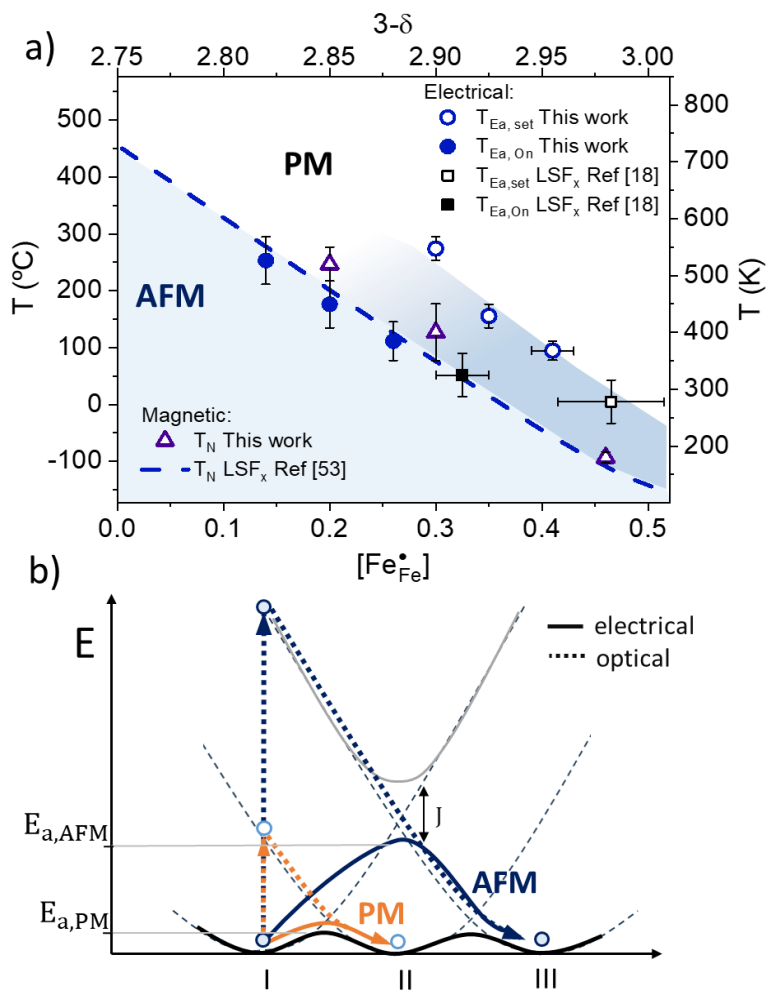

**Figure 6 a)** Magnetic phase diagram as a function of the concentration of holes. Electrical ($T_{E_a}$) and magnetic ($T_N$) transition temperatures in LSF50 are included. The figure shows the transitions from $E_a$=0.35eV ($T_{Ea,on}$) to an $E_a$=0.1eV ($T_{Ea,off}$), from both this study and Xie et al.[18] Neel Temperature ($T_N$) of LSF50 measured in this study and from Wattiaux et al.[53] are also included in the plot. **b)** Schematic representation of the origin of the change in $E_a$ and $E_{a,opt}$. Blue paths show the case of an AFM ordering of the spins in the material, where the hopping to the neighboring site is not allowed. Orange paths show the classical interpretation of polaron transitions to adjacent sites.

## 2.6. Co-LSF50 exchange modulation through magnetoionic

The analog control of electrical, magnetic, and optical properties of LSF50 thin films could be used in the future in emerging fields, such as neuromorphic computing and magneto-ionics. To demonstrate a possible application of the AFM/PM transition taking place in LSF50 when modifying its oxidation state, in this section we have investigated how this effect can be used to manipulate AFM/FM exchange coupling in Co/LSF50 bilayers. It is known that ferromagnetic (FM) layers can be influenced by adjacent AFM films through interfacial exchange interactions. This phenomenon either induces a shift in the magnetic hysteresis loop (i.e., exchange bias, EB), or an increase of coercivity ($H_C$) when the FM spins are capable of dragging the AFM interfacial spins, through a microscopic exchange torque that leads to an extra energy cost for the overall magnetization reversal. This last situation is typically observed for loops of FM/AFM bilayers measured close to $T_N$.[59,60]

To study this effect, a new sample with 100nm LSF50 was prepared following a procedure similar to the one proposed by Yildz et al. (see **Experimental section**).[61] The surface of the LSF50 film was proven (see **Supporting Information Section 1**), and the roughness was below 2 nm uniformly along the film, therefore excluding stress effects contributing to the $H_C$ variation. The sample allowed for an in-plane voltage application across two microfabricated gold rails forcing LSF50 to equilibrate at different oxidation states along the sample simultaneously from fully reduced to fully oxidized (see **Figure 7a**). A 2.5-nm Co layer and 2 nm Pt layer were sputtered on top of the LSF50 sample (see **Figure 7b**), and later studied by Magneto-Optical Kerr Effect (MOKE). The Co XAS spectra showed that no Co oxidation was visible in the as-deposited (AD) sample (see **Supporting Information Section 10**).

Hysteresis loops of the YSZ/LSF50/Co/Pt heterostructure were acquired at RT along the micro-fabricated rail, between the gold stripes using MOKE. **Figure 7c** shows the evolution of the hysteresis loops at X= 0, 0.75, 1.25 and 2.25 mm from the left contact of the rail. As can be seen, the AD Co loop in the reduced region (high $T_N$, left side of the figure) is less squared than in the oxidized region (low $T_N$, right side of the figure) and displays two contributions, with dissimilar coercivities. **Figure 7d** depicts the difference between the $H_C$ of the Co on top of the LSF50 and the Co on top of the gold stripes, which should remain unbiased, along the stripe ($\Delta H_C = H_{C,LSF50} - H_{C,gold}$). Interestingly, a clear critical point is evident in the middle of the rail, where the coercivity of the Co on top of the LSF50 at the right side of such point increases with respect to the coercivity on top of the gold. After the substantial increase, a modest decrease is also noticeable towards the rightmost area, but in no case the coercive field becomes the same as in the reduced part, which matches the coercivity values of the gold.

The sample was then subsequently field cooled (FC) at 50, 75, 100, 150, 200 and 250 °C, with 10 kOe applied to try to induce EB. The evolution of the loops after each FC are shown in **Figure 7c.** Eventually, the soft contribution seen in the reduced states disappears as the FC temperature increases and no loop shift was seen in any of the measurements. A general decrease in $H_C$ following each FC treatment was seen both in the Co located on top of the LSF50 and the gold. This phenomenon is ascribed to the loss of magnetic domains (and domain walls) in the Co layer, consequence of heating the sample and cooling it under a saturating field.[62] After the FC at 250 °C the loops in the oxidized part are drastically modified, likely due to Co oxidizing from the LSF50 underneath. **Figure 7d** tracks the evolution of the coercivity after each FC treatment.

The critical point of the increase of $H_C$ changed from X = 1.25 mm in the AD state to X = 1 mm after FC at 50 °C, where the coercivity increases by 32 and 28.2 Oe compared to the Co on top of the gold stripes, respectively. After the FC at 200 °C, $H_C$ majorly decreases independently of the location in the rail, suggesting already partial Co oxidation. This was later confirmed performing the FC at 250 °C, a process that modified not only the $H_C$ but also the shape of the loops in the regions with more oxygen abundance (see **Figure 7c**).

The $H_C$ trend observed in **Figure 7d** is consistent with the $T_N$ observations from the previous sections. The Neel temperature is supposed to vary from 700 K to 100 K from the most reduced state to the most oxidized state. In the middle of the rail, $T_N$ is presumably close to RT, allowing the FM Co spins to drag the interfacial AFM spins more easily, consequently increasing substantially $H_C$ in that region. As *X* further increases, $T_N$ decreases below RT and the material enters into the magnetic transition region. The FM layer remains pinned to the AFM. However, the Co requires less energy to drag the AFM spins, reducing the observed $H_C$.

This high-throughput design with the thin Co on top promises avenues for implementation in devices. Especially considering that without thermal treatments, the AFM LSF50 with $T_N$ close to RT already couples to the Co. Additionally, the possibility to tune magnetic and therefore, transport properties in an analog-like fashion, by gating with low voltages (<1 V), is appealing for memory devices and spintronics.

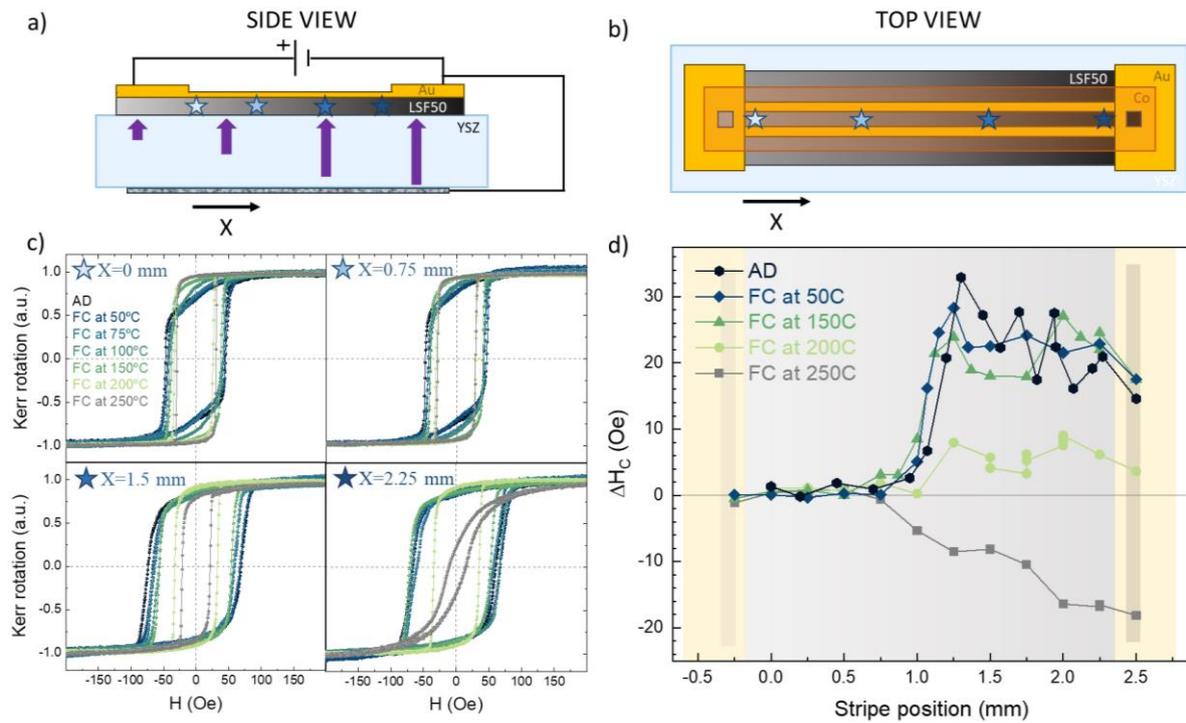

**Figure 7 a)** Schematic side view of the sample during the multi-state setting. b) Schematic top view of the sample after Co/Pt layer deposition. c) MOKE loops for the Co layer for 4 different positions along the rails measured in the as-deposited sample and after field cooling at different temperatures. d) Increase in coercivity of the Co loops after the different FC for the different positions along the rail compared to the Co deposited on top of the gold.

## 3. Conclusion

Fine-tuning of oxygen stoichiometry in mixed ionic-electronic conductive oxide of the La$_{1-x}$Sr$_x$FeO$_{3-\delta}$ family with x=0.5 was successfully achieved by applying small voltages through an electrochemical oxide-ion pump. Changes of several orders of magnitude in the polaronic conductivity were measured along oxygen deficiencies from δ=0 to 0.25 due to a variation of activation energy and pre-exponential factor. The origin of this electronic transition with temperature and point defect concentration was unveiled by simultaneously measuring of optical and magnetic properties. A close relation between the AFM/PM transition and the onset of the resistive state in LSFx was observed, with the $T_N$ strongly dependent on both $[Fe_{Fe}^{\bullet}]$ and temperature. Due to this magnetic transition, the polaron hopping varies from second-nearest-neighbors in the AFM state to the nearest-neighbor in the PM phase above $T_N$, giving rise to a decrease of the polaronic activation energy and an increase of electronic conductivity. The modulation of the defect concentration achieved by electrochemical potential reported here opens the possibility of using voltage to control the functional properties of mixed ionic-electronic conductor (MIEC) materials in an analog manner, enabling their application in switching and modulation devices as well as for their performance optimization. Finally, benefiting from the voltage-controlled AFM/PM transition, a magneto-ionic control of the coercivity in a coupled FM Cobalt layer grown on top of a LSF50 layer with different oxygen content was demonstrated.

## 4. Experimental Section/Methods

*Sample fabrication:* LSF50 thin films were fabricated using Pulsed Laser Deposition (PLD) on gadolinium-doped ceria (CGO)-coated YSZ (001) 1x1cm$^2$ substrates. The protective CGO layer with a thickness of approximately 20 nm was deposited onto the YSZ substrates to prevent the formation of secondary phases at the interface with LSF with YSZ.[63,64] A commercial pellet of CGO was used as target material, while the pellet of LSF50 were prepared by solid state synthesis as explained in Tang et al, 2021.[36] All the layers were deposited employing a large-area system from PVD products (PLD-5000) equipped with a KrF-248 nm excimer laser from Lambda Physik (COMPex PRO 205). The films were grown with an energy fluency of 0.8 J cm$^{-2}$ per pulse at a frequency of 10 Hz. The substrate was kept at 600 °C, in an oxygen partial pressure of 0.0067 mbar during the deposition and the substrate-target distance was set to 90 mm. The deposition of the functional layer was done using a microfabricated Si mask with a size of 2.5x3.5mm$^2$ centered in the substrate. The thickness and roughness of the deposition were checked prior to the following fabrication steps allowing for a measurement of 99nm thicknesses for LSF50 (see **Supporting Information Figure S1**). To electrically contact the functional thin film, a conductive 10nmTi/100nmAu electrode was microfabricated by photolithography and thermal evaporation. The thin Ti layer in the contact is deposited in order to enhance a good adhesion of the Au to the sample. To prevent the equilibrium of the functional layer with the atmosphere when going to higher temperatures, a last capping layer of alumina (Al$_2$O$_3$) of approximately 100nm was also deposited via PLD (see a schematic of the samples in **Figure 1a**). Alumina was chosen as the capping material because of its low oxygen mobility and negligible oxygen stoichiometry variation in a wide range of oxygen partial pressure ($pO_2$) and temperatures. Another important property of the alumina capping is its low absorption, which allow to measure the optical properties of the functional layer with more precision. Silver paste was also used on the backside of the YSZ substrate as counter electrode.

For the MOKE characterization of Co-LSF50 interaction, a different sample was used. In this case, 100nm LSF50 deposited on 20 nm gadolinium-doped ceria (CGO)-coated Yttria Stabilized Zirconia (YSZ) (001) 0.5x0.5 cm$^2$ substrates was prepared. For this sample, 10 nm Ti/100 nm Au with a design consisting in two pads connected by a pair of parallel stripes was used as working electrode (WE) (see sketch in **Figure 7a and 7b**). The separation between pads was 2.3mm and the stripes had 50 µm width and 100µm separation between them. One of the top pads was grounded with the silver painted in the bottom part serving as counter electrode (CE). This allowed for an in-plane voltage application that also forced the voltage difference between the WE and the CE. The linear potential drop in the layer allowed for the equilibration of the functional layer's oxidation state to different potentials all along the stripes' direction, i.e., allowing to have all the studied $[Fe_{Fe}^{\bullet}]$ in a single sample simultaneously (see **Figure 7a**). After the state was set at 350ºC applying a voltage difference between pads of -0.55V, the states were quenched by cooling down at 30ºC/min while keeping the voltage difference. Color gradient of the LSF50 in the stripe's direction was observed, confirming the multi-state presence. A 2.5-nm Co layer and 2 nm Pt layer were sputtered on top of the previously described microfabricated gold rails in order to study the imprint of the antiferromagnetic domains of LSF50 onto the Co layer.

*Thin Film Characterization:* X-ray diffraction (XRD) was recorded using a Bruker D8 Advanced diffractometer with Cu Kα radiation (λ = 1.5406 Å) in a coupled Θ–2Θ Bragg–Brentano configuration. X'Pert PRO MPD analytical diffractometer (Panalytical) at 45 kV and 40 mA using CuKα radiation was also used for a deeper microstructural characterization and phase identification. The topography of the thin films and film thickness measurements were also assessed with an atomic force microscope XE 100 model from Park System Corp in non-contact mode.

*Electrical and electrochemical characterization:* Electrical conductivity measurements of the thin film were performed at different temperatures using a heating stage (Linkam instruments THMS600E-2) and 4-probe measurements with a Kithley 2400. Electrochemical characterization of the sample was also performed using a Biologic (model SP-150). A DC voltage bias from 0.1 V to −0.45 V was applied at different bias steps between the top gold contacts and the back-side Ag counter-electrode. Once the current was stabilized, the electrochemical impedance spectra (EIS) were recorded within a frequency range of 0.1 Hz-1MHz and an amplitude of 10mV. After the stabilization, electronic conductivity of the LSF thin film was measured. In order to rule out the resistance of the contacts (e.g. from Ti interlayer oxidation), the film conductivity (σ) was measured using a 4-point probes method, considering the distance between the voltage sense connections (L=1.5mm), the width of the layer (d=3.5mm) and its thickness (t). Conductivity values were obtained through the relation between the electrical and geometrical parameters given by $R=I_R/V_R=L \cdot t^{-1} \cdot d^{-1} \cdot \sigma^{-1}$. All the experiments were carried out in atmospheric air.

*Spectroscopic Ellipsometry Measurements:* Spectroscopic ellipsometry, performed with the UVISEL ellipsometer from Horiba scientific, measured the optical constants of LSF thin films within a photon energy range from 0.6 to 5.0 eV with 0.05 eV intervals, using a 70° incident light beam. The data acquired from ellipsometry were subjected to modeling and fitting via

Horiba Scientific's DeltaPsi2 software. In-situ ellipsometry measurements were carried out at different temperatures, following each voltage bias step.

*Magnetization characterization:* In-plane temperature vs Magnetization measurements were performed on a LakeShore 8600 series vibrating sample magnetometer (VSM) by means of an adjustable oven. The field was kept constant at 1kOe during the measurements. Cooled nitrogen (<350 K) and heated argon (>350 K) were injected in the oven to ensure thermal conductivity inside it. A NanoMOKE3 magnetooptical Kerr microscope was used to perform longitudinal MOKE measurements with a linearly polarized laser spot size of 3 μm and a wavelength of 660 nm. An in-plane variable magnetic field was applied along the gold stripes with a maximum value of ±400 Oe during the measurements, with a frequency of 7.3 Hz. To account for statistics at this high frequency, a minimum of 300 loops were always averaged before obtaining the shown loops.

*PEEM imaging*: X-ray photoemission-electron microscopy studies were carried out using the SPEEM end-station at UE49-PGMa beamline at the synchrotron radiation facility BESSY II operated by Helmholtz- Zentrum Berlin (HZB)[65]. The angle of incidence of the incoming X-ray radiation with respect to the sample surface was 16°. Magnetic imaging was always performed in zero external field. PEEM images were collected at 721.0 and 722.5 eV, at the Fe $L_2$-edge. A total of 80 images, each with a 3s integration time, were collected at each energy. Field of view was 10μm x 10μm, 514 x 514 pixels. Prior averaging the images, their drift was corrected. The XMLD images were obtained as (A –B)/(A + B), where A and B were the averaged images for each of the chosen energies, respectively. Due to the low XMLD signal, a modulation of the intensity caused by interferences of the incoming beam with the exit slit is visible across the original images. This modulation was removed by subtracting to the XMLD image a twin image after applying a 50 X 50 averaged filter.

**Supporting Information**

Supporting Information is available from the Wiley Online Library or from the author.

**Acknowledgements**

This project received funding from the European Union's Horizon 2020 research and innovation program under grant agreement No. 824072 (HARVESTORE), No 101066321 (TRANSIONICS) and, No. 101017709 (EPISTORE) and under the Marie Skłodowska-Curie grant agreement No 840787 (Thin-CATALYzER). F.C. acknowledges funding from a Marie Skłodowska Curie Actions Postdoctoral Fellowship grant (101107093). The authors acknowledge support from the Generalitat de Catalunya (2021-SGR-00750, NANOEN). Financial support from the European Research Council (2021-ERC-Advanced REMINDS Grant Nº 101054687), the Spanish Government (PID2020-116844RB-C21) and the Generalitat de Catalunya (2021-SGR-00651) is acknowledged. The authors thank the Helmholtz-Zentrum Berlin for the provision of access to synchrotron radiation facilities and allocation of synchrotron radiation at SPEEM end-station of UE46-PGMa.

Supporting Information

**Analog control of electrical conductivity in La$_{0.5}$Sr$_{0.5}$FeO$_{3-\delta}$ through oxygen deficiency induced magnetic transition**

*Paul Nizet, Francesco Chiabrera\*, Nicolau López-Pintó, Nerea Alayo, Philipp Langner, Sergio Valencia, Arantxa Fraile-Rodriguez, Federico Baiutti, Alevtina Smekhova, Alex Morata, Jordi Sort, Albert Tarancón\**

**SI1: Surface characterization**

Atomic Force Microscopy (AFM) surface analysis of the deposited films was done prior to the Al$_2$O$_3$ capping. **Figure S1a** shows the surface roughness of LSF50. As it can be seen, LSF50 surface present a smooth surface with roughness under 2nm. The same stands for the Al2O3 capping surface presented in **Figure S1b**.

To corroborate the thickness previously measured with Spectroscopic ellipsometry of each layer, a measurement on the edge of the deposition was performed. **Figure S1c** shows the step-like edge of the deposition of LSF50 which confirmed its 99nm thickness. The measured thickness from AFM was later used both for electronical conductivity calculations and optical properties fitting.

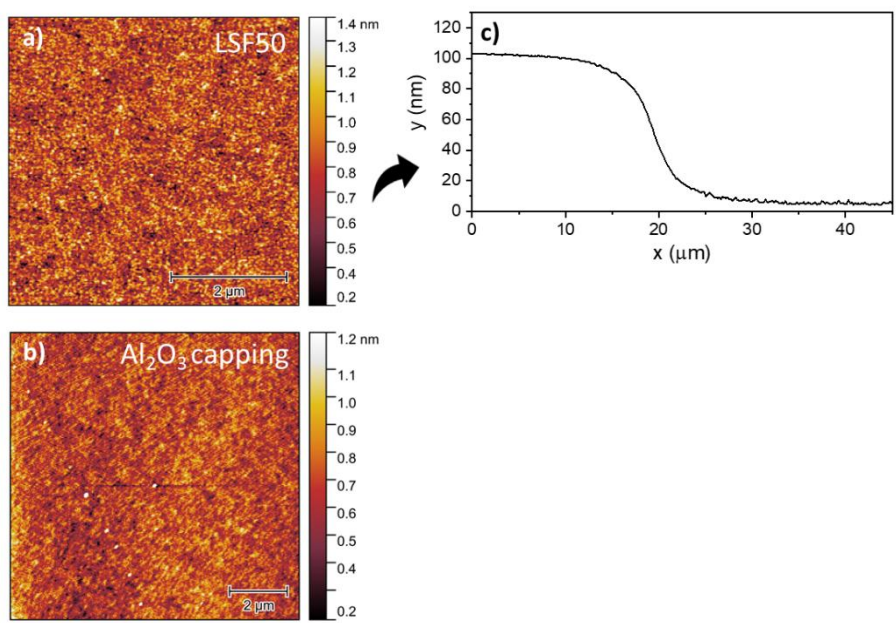

**Figure S1** AFM images of the as-deposited **a)** LSF50 and, **b)** Al2O3 capping. **c)** AFM line scan of the edge of LSF50 thin film deposited layer.

**SI2: Crystallographic characterization**

As part of the initial sample characterization and property change investigation, room temperature XRD measurements were conducted. **Figure S2a** shows the as prepared sample diffraction spectrum. A clear (001)-orientation of the CGO on top of the YSZ can be seen which also favors the preferential (001)-orientated growth of the LSF50 although a secondary (011)-LSF50 peak is also seen. Phi scans of the (202)-CGO and (101)-LSF peaks were performed to confirm the epitaxial growth of the layers (**Figure S2b)**. An in plane 45º rotation of the cell of the LSF50 with respect to the ones of CGO and YSZ is seen, as expected from their cell parameter values. **Figure S2c** shows the pole figure at the (111)-LSF50 reflection. In the figure, clear peaks appear each 90º at an angle of 46º, corresponding to the (001)-orientation. Nevertheless, not clear (011)-orientation contribution could be seen with a 45º shift with respect to the (001)-peaks centered at around 35º. This indicates that the film main orientation is in the (00l) direction.

In conclusion, room temperature XRD of the as prepared sample shows a good epitaxial growth of the functional LSF50 film with a preferred (l00) orientation. The absence of these BM peaks for LSF50 at various oxidation states further confirms that no long-range ordering of vacancies is present in the material. There is no emergence of half order peaks ($\frac{1}{2}$ 00 and $\frac{5}{2}$ 00) i.e., it consistently remains in the PV phase.

Furthermore, a trend in the peak maxima of the 100 and 200 peaks can be observed upon oxidation/reduction of the material, see **Figure S1d**. **Figure S1e** illustrates the calculated unit cell parameter for each spectrum obtained as a function of $[Fe^{4+}]$. The values of the LSF50 sample align with the observed trend concerning $[Fe^{4+}]$ for different LSFx with distinct x.[1,2] Also, it's increase in the cell parameter is also in agreement with the variation of Fe-O bond length predicted by a simple hard sphere approximation as the smaller $Fe^{4+}$ (58.5 pm) are substituted by the larger $Fe^{3+}$ ions (64.5 pm).[3,4]

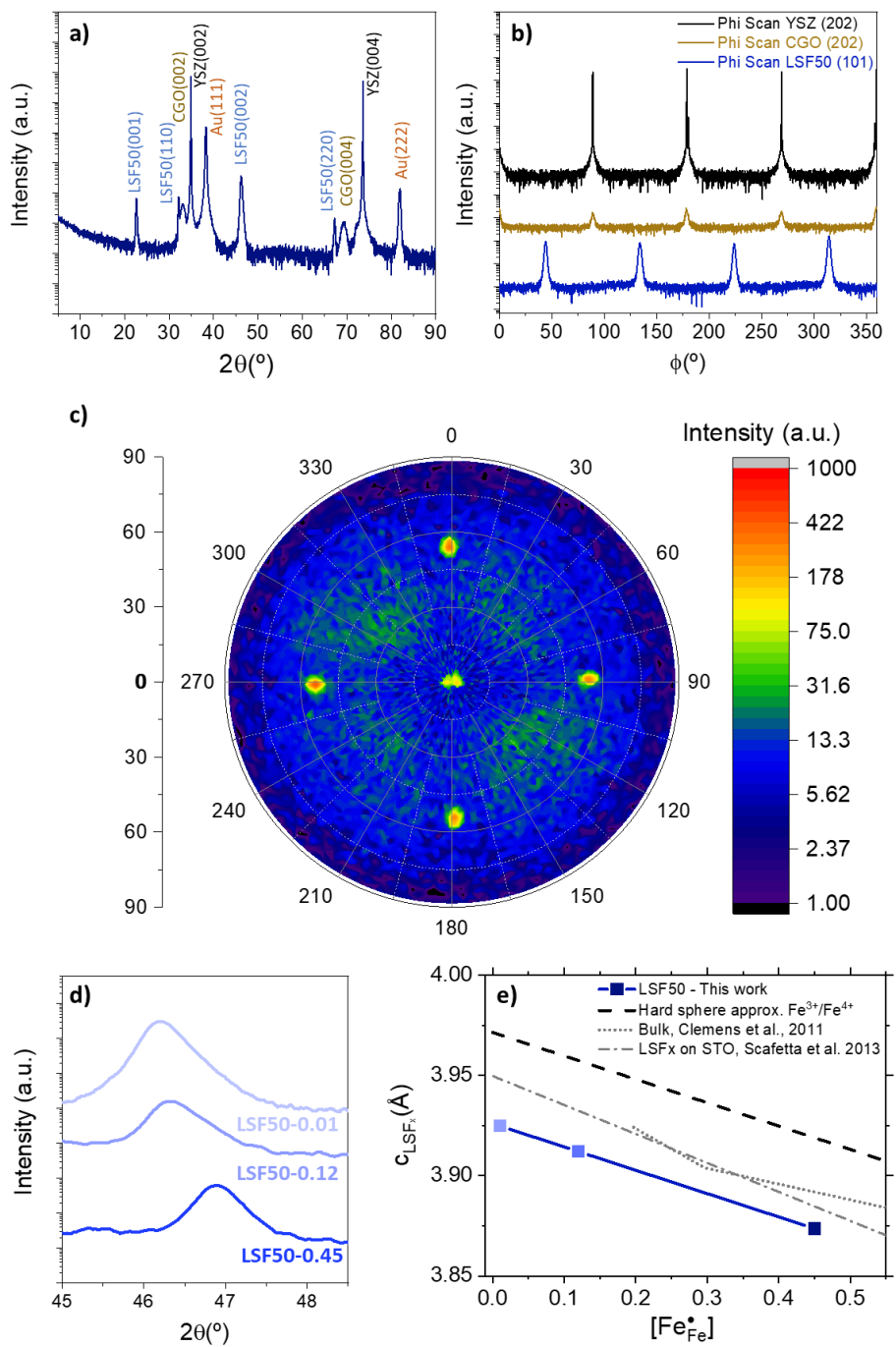

**Figure S2** a) XRD pattern of the as-deposited (LSF50-AD) layer. b) Phi scans of the YSZ(202), CGO(202) and LSF50(101) for the as prepared sample. c) Pole figure at the LSF(111) reflection. d) XRD pattern of the 002-diffraction peak for the voltage-reduced/oxidized LSF50 indicated with the calculated $[Fe'_{Fe}]$. e) Room temperature XRD pattern of the LSF50 samples after the application of different $V_G$ at 350ºC indicated by the calculated $[Fe^{\bullet}_{Fe}]$. Reference measurements on LSFx cell parameters from Clemens et al. and Scafetta et al. are also shown[1,2]

**SI3: Sample characteristics and oxidation state vs V$_G$ (pO2)**

As well explained in the main text, the sample configuration allows for oxygen incorporation into the functional layer exclusively through the YSZ and the alumina capping is used to block the direct oxygen entry from the atmosphere. The behavior of the functional layer is expected to be the one proposed by Maier et al. for a surface-limited incorporation mechanism.[1] For a system limited by the insertion of the mobile ions and not the diffusion inside the layer, the behavior of the layer can be approximated to a capacitor. **Figure S3a** illustrates the Nyquist impedance plot for one of the samples measured between the gold contacts and the silver counter electrode (i.e., across the YSZ substrate). A semicircle is clearly seen in the high-frequency region due to mobile charge motion in the YSZ electrolyte. Taking into consideration the dimensions of the deposited layer, a conductivity of 20 μS·cm $^{-1}$ was calculated, which is in perfect accordance with YSZ conductivity at 350ºC. Also, the low-frequency region shows a capacitive behavior with a constant Real Impedance contribution but an increase in the Imaginary Impedance part as the frequency is lowered. This corroborates again the good capping behavior of the Al2O3 layer, as the oxygen inserted/extracted from the functional LSFx layer cannot flow through the capping layer and is only being accumulated in the LSFx layer. Moreover, since no DC current is measured under the application of an electrochemical bias, all the charge that flows through the system is indeed the oxygen inserted/extracted into the LSFx layer, which can be determined through current integration.

$$Q(t) = \int_0^t I(t)dt \qquad (S1)$$

Being $Q(t)$ the charge incorporated in the system i.e., twice the number of incorporated oxygens. **Figure S3b** illustrates the current behavior when applying voltage differences (ΔV) relative to the counter electrode, ranging from -0.4V to 0.1V. With the integration of Equation S1 allowing us to ascertain the oxygen quantity within the film as it attains equilibrium at each applied voltage.

It is understood that the process of oxygen incorporation within the LSFx layer adheres to the following relationship.

$$\frac{1}{2}O_2 + V_O^{\cdot\cdot} + 2Fe_{Fe}^x \leftrightarrow O_O^x + 2Fe_{Fe}^{\cdot} \qquad (S2)$$

Where $Fe_{Fe}^{\cdot}$, $O_O^x$, $V_O^{\cdot\cdot}$, and $Fe_{Fe}^x$ denote the $Fe^{3+}$, oxygen ions, oxygen vacancies and $Fe^{4+}$ holes species respectively according to the Kroger-Vink notation. Considering that, within the used range of temperature and partial oxygen pressure, the presence of other iron oxidation states, such as $Fe^{2+}$ electrons and $Fe^{5+}$, is negligible.[5–7] Consequently, the material's charge neutrality and side balance equations will be controlled by

$$[Sr'_{La}] = 2[V_O^{\cdot\cdot}] + [Fe_{Fe}^{\cdot}] \quad (S3)$$

$$[La_{La}^x] + [Sr'_{La}] = 1 \quad (S4)$$

$$[Fe'_{Fe}] + [Fe_{Fe}^x] + [Fe_{Fe}^{\cdot}] = 1 \quad (S5)$$

$$[O_O^x] + [V_O^{\cdot\cdot}] = 3 \quad (S6)$$

Here; $[Fe_{Fe}^x]$, $[Fe_{Fe}^{\cdot}]$, $[O_O^x]$, $[V_O^{\cdot\cdot}]$, $[La_{La}^x]$, and $[Sr'_{La}]$ denote the concentration per unit cell of $Fe^{3+}$, $Fe^{4+}$ holes, oxygen ions, oxygen vacancies, lanthanum, and strontium respectively. To determine the oxidation state, the procedure followed was to reduce the layer by applying negative voltages until it could no longer be reduced i.e., when no additional current flowed despite applying increasingly negative voltages. This ensured that all Fe atoms in the sample were in the $Fe^{3+}$ state. From there, the sample was stepwise oxidized by returning to higher voltages. Following Equation S3 and S6, the integration of current, indicating the quantity of oxygen incorporated into the layer, was thus directly proportional to the quantity of $Fe^{4+}$ ions.

Various studies have debated on the underlying mechanism that triggers the ionic motion, and hence, the oxidation state alteration of these materials when subjected to applied voltages. It has been widely acknowledged that the voltages applied across the LSF layer (working electrode) and the silver counter electrode induce changes in the chemical potential of oxygen. This, in turn, leads to an associated change in the equivalent partial pressure of oxygen in line with the Nernst potential.

$$p_{O_2}^{LSF} = p_{O_2}^{Ref} \cdot exp\left(\frac{4e\Delta V}{k_b T}\right) \quad (S7)$$

where $p_{O_2}$ stands for oxygen partial pressure and superscript indicates if it is referred to the *LSF* equivalent one or the environment reference $p_{O_2}$. *T* is temperature of the sample, $k_b$ is the Boltzmann constant, *e* is the electron charge, and *ΔV* is the potential difference between the silver counter electrode and the LSF thin film. This change in the $p_{O_2}^{LSF}$ is the responsible of changing the equilibrium of the material, making it oxidize or reduce depending on its defect chemistry.[8]

To demonstrate that the final oxidation state is the same after the application of a certain *ΔV* to the film, reversibility of the oxygen (dis)charging from the film was proven. **Figure S3c** shows the experimental conductivity of the LSF50 film after different applied V$_G$=*ΔV*. Three different states were chosen and set following a random order. Conductivity was measured after current stabilization following the two-step procedure explained in the main text. The material shows good cyclability of its electronic conductivity properties and no hysteretic behavior was observed.

Measurements to prove the stability of the state after the application of a $V_G$ were also conducted. **Figure S4a** shows the voltage application followed by the subsequent open circuit voltage measurement (OCV). The measurement shows a good retention of the OCV=-0.1V for up to 10 hours after the application of the same gating voltage to change the oxidation state $V_G$=-0.1V. Temperature retention of the state was also checked by performing three cycles of 200-350ºC temperature variation. **Figure S4b** shows the zoom in on the temperature cycles. When temperature goes down, OCV is also lowered. This is well explained by the defect chemistry of the material and the Nerst potential (Equation S7), as lower temperatures translate into greater potentials $\Delta V$. For temperatures lower than 250ºC, the YSZ resistance becomes too high for the OCV to be measured with precision. This leads to a higher error in the signal as seen in **Figure S4a**, nevertheless, a clear trend can be seen following the blue lines presented in **Figure S4b**. After each temperature cycle, an OCV of -0.1V is recovered meaning that the state is conserved even if T is changed.

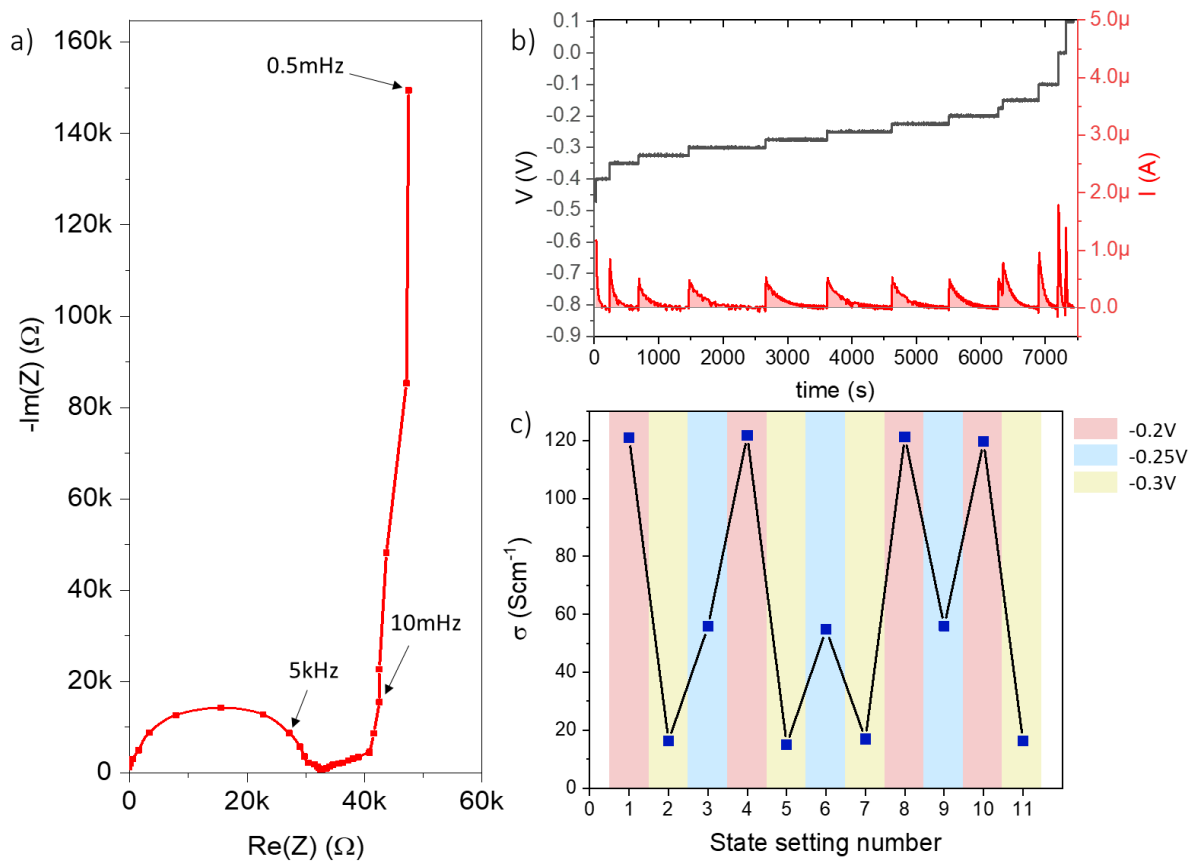

**Figure S3 a)** Nyquist impedance plot performed in the LSF50 sample. **b)** Applied writing bias ($V_G$) and measured current vs time for the LSF50 sample at T=350C used for current integration to obtain the oxidation state vs $pO_2$. **c)** Conductivity of LSF50 at T=350C after the application of different $V_G$ in a randomly chosen order, proving that the equilibrium state only depends on the chosen $V_G$ (equivalent $pO_2$).

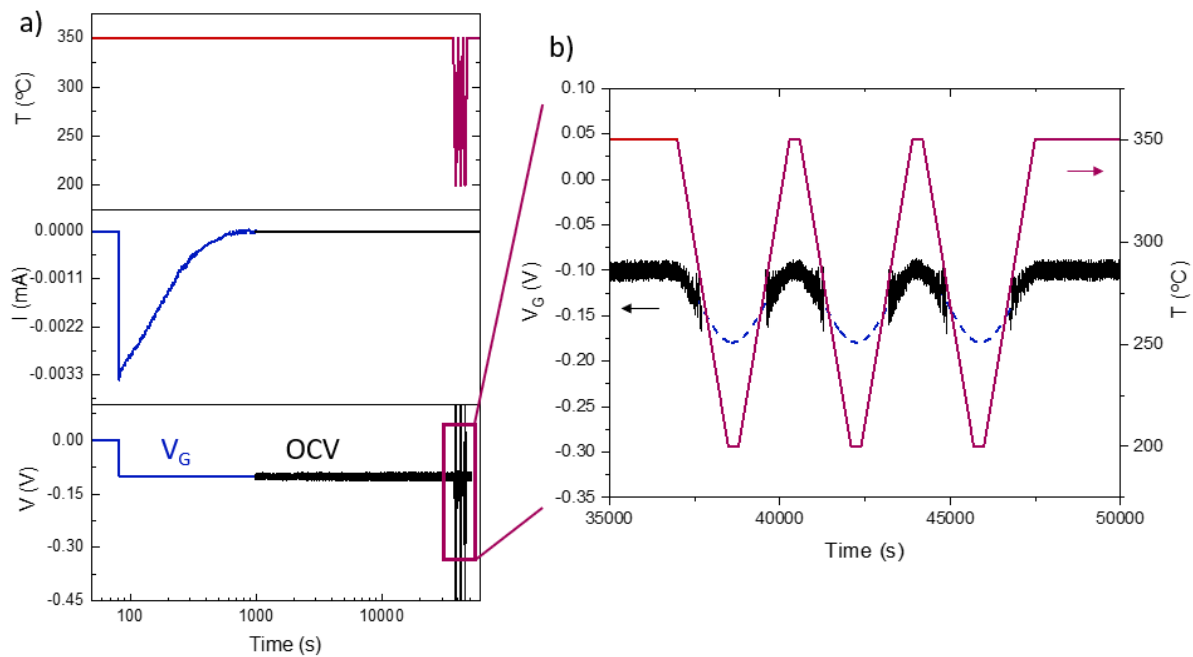

**Figure S4** a) State retention and capping retention study showing the initial setting of the state, via the application of a $V_G$=-0.1V until $I_G$=0. OCV was measured during 10h before the application of 3 thermal cycles to show that no loss of the OCV was present after more than 13h. b) Thermal cycles showing the retention of the OCV when going back to 350C. Blue dotted line shows the missing data due to the error in OCV readout at lower temperatures.

**SI4: Optical properties extraction from ellipsometry measurements and analysis**

The ellipsometry raw data are modeled and fitted using the DeltaPsi2 software. The dielectric properties of LSF50 thin film are obtained using 4 Lorentzian oscillators following the formula:

$$\varepsilon = \varepsilon_o + \sum_{j=1}^{n} \frac{f_j \omega_{0,j}^2}{\omega_{0,j}^2 - \omega^2 + i\gamma_j \omega} \qquad (S8)$$

Here, $\varepsilon_o$ is the high frequency dielectric constant, $\omega$ is the angular frequency of the light beam, $\omega_{0,j}$, $f_j$, $\gamma_j$ are the resonance frequency, intensity, and the broadening of each oscillator. This total dielectric constant is used to obtain the n and k of the material that fits best the $\Psi$ and $\Delta$ spectra.

The inset of **Figure SI5** shows the model used for the fitting. It consists in 5 layers with specified thickness: the one side polished YSZ substrate, the CGO interlayer, the LSF50 thin film, the Al2O3 capping and a top layer consisting of 50% vol. of Void and 50% vol. of Al2O3 thin film, representing the surface roughness of the sample). The thickness of each layer was previously measured by AFM and so is kept constant during the fitting. Also, optical properties of the substrate, CGO and Al2O3 were measured beforehand as a function of temperature. In this way, the fitting is reduced to the parameters of the oscillators of the functional material. The errors of the optical conductivity presented in this work were calculated by the sensitivity of the parameters of the corresponding oscillator given by the DeltaPsi2 software during the data fitting procedure. The resulting refractive index n and extinction coefficient k obtained for each of the studied materials as a function of temperature and oxidation state is plotted in **Figure S5**, as well as the optical conductivity represented by the formula:

$$\sigma(\omega) = \frac{4\pi n k}{\lambda Z_o} \qquad (S9)$$

With $Z_o = \sqrt{\mu_o/\varepsilon_o} = 377\ \Omega$ being the impedance of free space. This optical conductivity can be interpreted as the electrical conductivity in the presence of an alternating electric field of frequency ω.

Similar trends as the ones commented on the main text can be seen to be followed at all the studied temperatures. Notably, the isosbestic point observed for LSF50 in the optical conductivity can also be seen in the extinction coefficient k.

As a matter of example, both the intensity and energy of transition A can be subject to analysis. **Figure S6** illustrates a linear correlation between the intensity of the transition and $[Fe^{4+}]$ hole concentration, consistent with prior findings by Tang et al. in LSF50.[8] From the Fermi Golden Rule, we know that the probability of the transition to take place is proportional to the density of states (DOS) of the final level. As this transition corresponds to a transition to the hole-induced band (see **Figure S13** for the detailed explanation of the band evolution), the DOS of the final state will be proportional to the concentration of holes i.e., proportional to $[Fe^{4+}]$. The verification of this linearity confirms that the $[Fe^{4+}]$ measured with current integration is indeed the one present in the sample.

The A-transition peak analysis for LSF50 is shown in **Figure S6b**. The results show both the linear relation of the intensity with $Fe^{4+}$ hole concentration and the energy of the maximum for that transition, that follows the behavior measured for the Ea also shown in the main text. Temperature dependance on optical properties for a given oxidation state can also be studied. **Figure S7a** shows the optical conductivity of three different states and their respective first peak optical conductivity. The analysis of the entire spectrum (**Figure SI7a**) reveals that the temperature variation primarily affects transitions A and C. A general trend can be observed as both transitions decrease with increasing temperature. The constancy of transition B with temperature can be attributed to its nature as a direct transition.[9,10] **Figure S7b** examines the variation of transition A. In the most oxidized case, $[Fe^{4+}]$= 0.5, the shape and position of the peak barely change and only the intensity of the maximum decreases, while for intermediate states ($[Fe^{4+}]$= 0.26 and $[Fe^{4+}]$= 0.2) a progressive decrease of peak energy is observed as T is increased, in agreement with the measured change in Ea. Significant alterations in energy bands due to the expansion/contraction of the unit cell with temperature is not expected.[11] Those reasons lead us to believe that this temperature-induced change in the optical transition to lower energies, which is also observed as a change in the conductivity activation energy ($E_a$), must be attributed to a prohibited electronic transition, which becomes less constrained with the increase in thermal energy.

The Band gap change was also calculated using the Tauc plot corresponding to direct-forbidden transitions[9,10] as plotted in **Figure S8**. The results are shown in **Figure S8d** and **Figure S9** demonstrate the closing of the gap when going to more oxidized states. Notably, the gap is smaller for higher temperatures as expected from the Varshni effect.[12] The Energy of the gap ($E_{gap}$) measured for LSF50 is compared with a reference measuring the same parameter for LSFx with different Sr content after oxidation them[9] as shown in **Figure S9**. The results show that the trend of the reference is followed for the results measured in the PV stable phases.

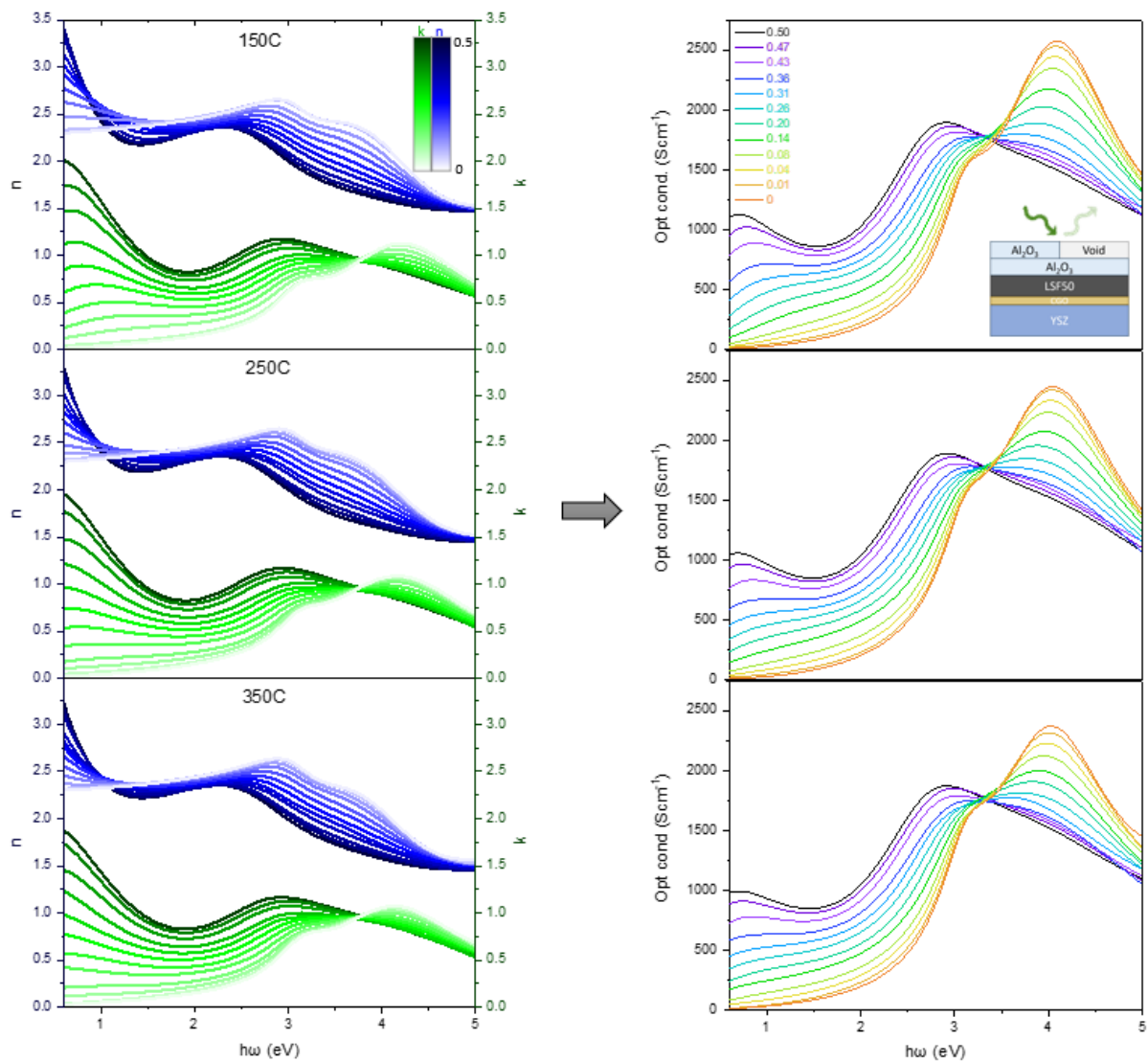

**Figure S5.** Figure showing the obtained n and k for the different oxidation states for LSF50 indicated by the $[Fe^{4+}]$ at different temperatures (left panel) and the corresponding calculated optical conductivity (right panel). The inset illustrates the model used to fit the optical properties.

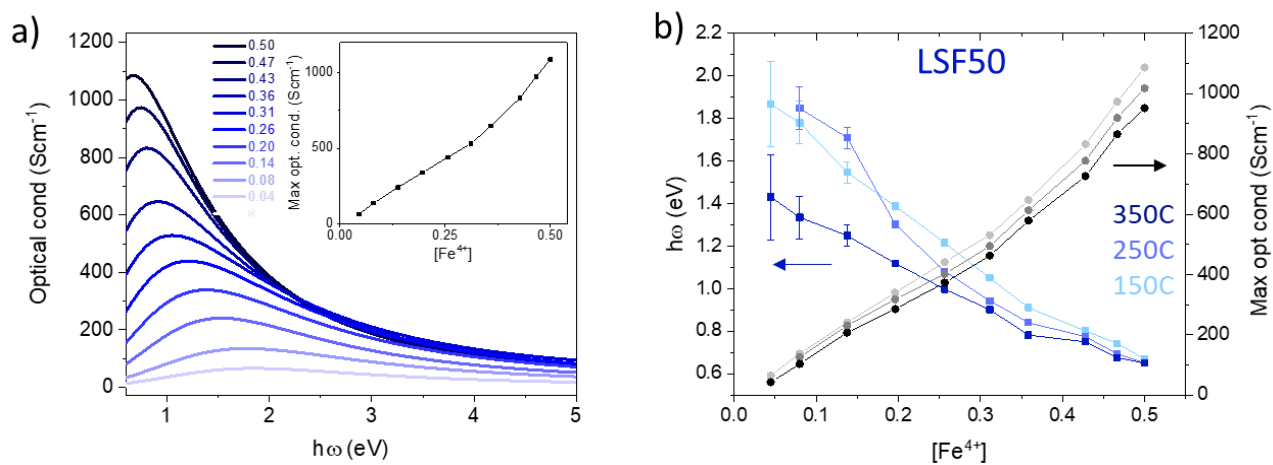

**Figure S6. a)** Analysis of the Transition A for LSF50 at 150C. Inset figure shows the linearity of the maximum of the peak as a function of the oxidation state ($[Fe^{4+}]$). **b)** Phonon energy of the maximum of the optical conductivity and its value as a function of the oxidation state, noted as $[Fe^{4+}]$ and temperature.

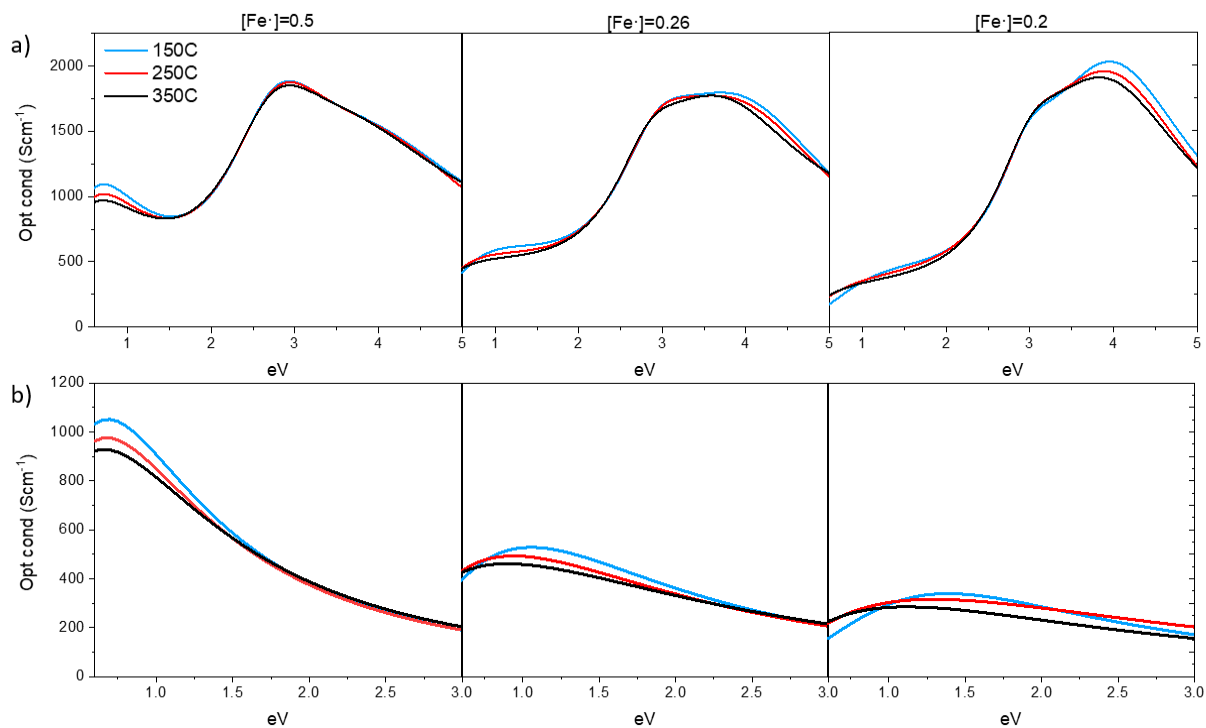

**Figure S7.** Optical conductivity of LSF50 **(a)** and first transition peak **(b)** at three different oxidation states indicated by the $[Fe^{4+}]$ measured at different temperatures.

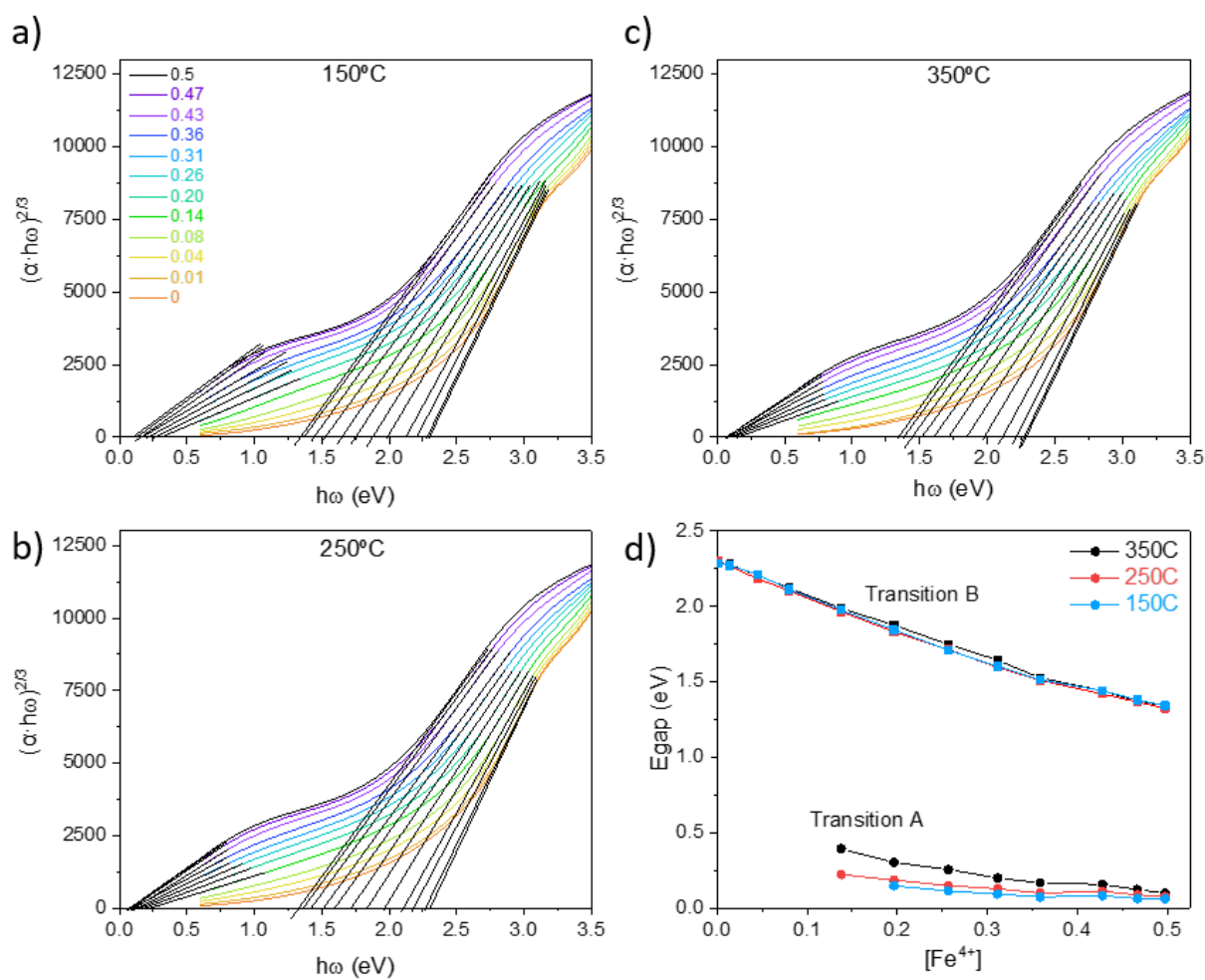

**Figure S8.** Tauc plots for direct-forbidden transitions for the LSF50 data measured at different temperatures **(a-c)** and gap energy measured from them **(d)**.

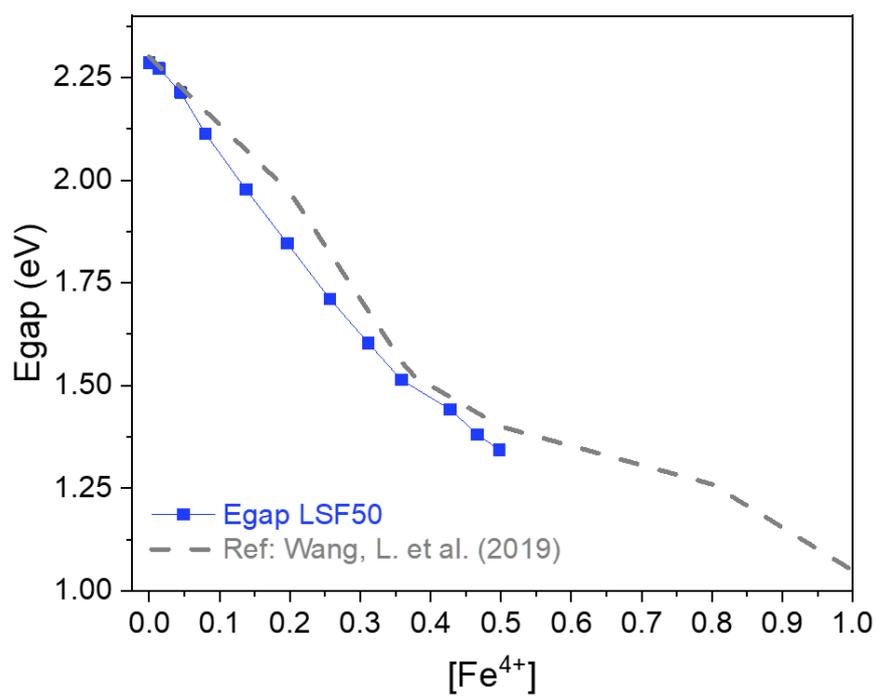

**Figure S9.** Comparison of the $E_{gap}$ depending on the oxidation state with the data of different oxidized LSFx from Wang et al.[9]

**SI5: Non-adiabatic and adiabatic polaron hopping comparison**

As explained in the main text, conductivity measurements could be well fitted by both the adiabatic and the non-adiabatic polaron hopping models. **Figure S10a** shows the Arrhenius plot for the non-adiabatic polaron hopping model for the LSF50 data for the different oxidation states measured indicated by the $[Fe^{4+}]$. The figure shows a good linearity as expected from the formula:

$$\sigma(T) = \frac{\sigma_o}{T^{3/2}} exp\left(\frac{E_a}{k_b T}\right) \tag{S15}$$

As in the adiabatic hopping formula, $E_a$ refers to the activation energy of the hopping, $k_b$ is the Boltzmann constant and $\sigma_o$ is the pre-exponential factor. Different references prefer to use the non-adiabatic[13] over the adiabatic model and vice versa,[14,15] although some comment that the data can also be fitted by both models.[13] In our case the decision on the adiabatic model was taken after observation of the optical properties. From the discussion given in the main text, one could see that $E_a$ could be estimated from the maximum absorption of the optical transition to the hole induced level ($h\omega$) as

$$E_{a,opt} = \frac{1}{4} h\omega - J \tag{S16}$$

Where $J$ was the transfer energy. In the non-adiabatic model, $J$ can be neglected and so the optical transition is four times the activation energy of the hopping. This last approximation does not match well with the experimental data as there is still a ~0.1eV factor to make both values agree.

For more specific analysis, as happens in the main text, the comparison between $E_a$ and the preexponential factor of the hopping can give a wide vision on the phenomena underneath the variation of conduction mechanism. **Figure S10b and S10c** shows the variation of these two parameters when applying the non-Adiabatic model, leading to the same behavior and conclusion presented in the main text.

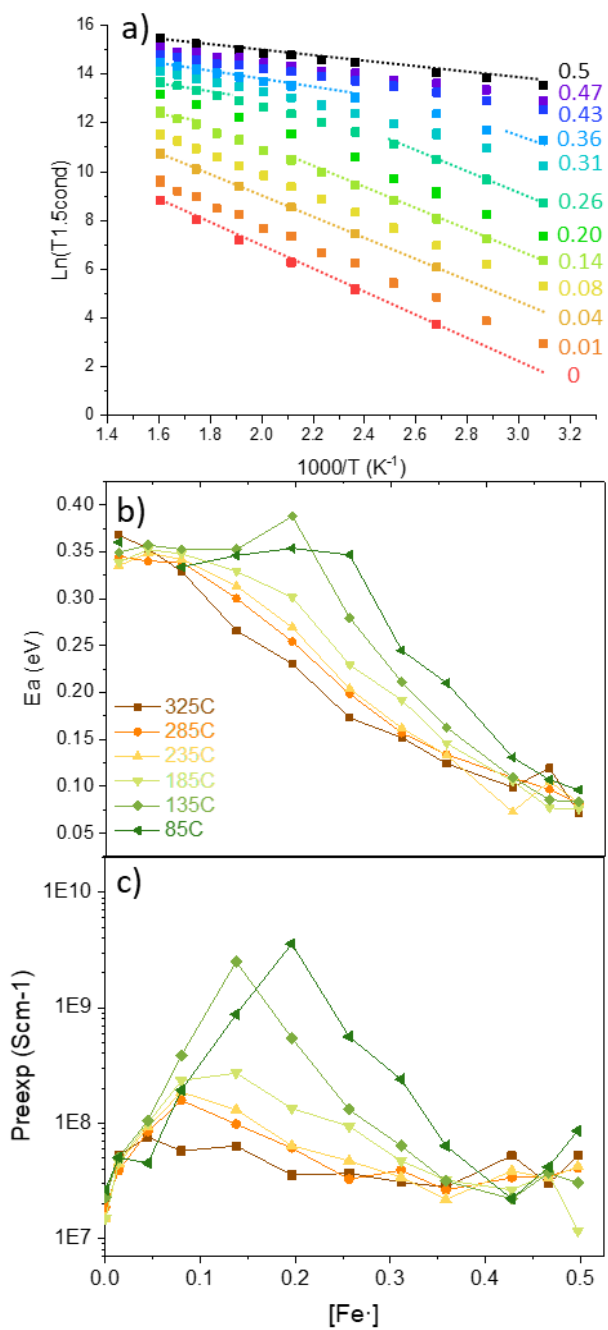

**Figure S10 a)** Arrhenius plot for a non-adiabatic hopping for the experimentally measured LSF50 conductivity at different oxidation states indicated by different colors and the corresponding $[Fe^{4+}]$. **b-c)** Study of the parameters of the hopping for LSF50. The figure shows the variation of $E_a$ (b) and the preexponential factor (c) for the non-adiabatic hopping model.

**SI6: Activation energy transition temperature ($T_{Ea}$) calculi and conductivity behavior comparison**

To study the origin of the $E_a$ change with temperature and oxidation state, **Figures SI11a and SI11b** present the data derived from the Arrhenius in **Figure 3b** for the activation energy and pre-exponential term measured for LSF50. For comparison, data from stoichiometrically oxidized LSFx at lower temperatures were included temperatures.[13] The processing of these latter data involved dividing the measured temperature range in their study into three parts, calculating the activation energy and corresponding pre-exponential term for adiabatic hopping. As these additional data were obtained from differently doped LSFx samples, the resulting values are plotted as a function of the expected $[Fe^{4+}]$ in **Figure S11**.

As observed, both characteristics exhibit a pattern akin to the LSF50 data measured in this paper. At lower temperatures, there is a constant $E_a$ coupled with an exponential rise in the pre-exponential factor, and as $[Fe^{4+}]$ approaches higher values a decrease in $E_a$ is observed and the trend in the pre-exponential is also lost. This affirms that the observed behavior in LSF50 measurements is not directly linked to the variation in $[V_o^{..}]$ but is rather attributed to the change in $[Fe^{4+}]$ holes resulting from charge compensation. To pinpoint the $E_a$ transition points, a threshold value for the high $E_a$ state was defined as $E_a$=0.32 eV. Consequently, the transition points ($T_{Ea, on}$ and, $T_{Ea, set}$) are identified as the first temperature where $E_a$ falls below this threshold and the one in which it first has an $E_a$< 0.125 eV. To have greater accuracy in the temperature of the transition, **Figure S12** shows the $E_a$ measured at 8 temperature ranges (instead of 6 as in the main manuscript) and shows the data from Xie et al.[13] Green and blue areas in the graph show the values of $E_a$ that are above and under the transition range, respectively. Three points for each threshold are directly obtained from our data analysis corresponding to the three $[Fe^{4+}]$ that have $E_a$ values in both regions and a fourth point arises from the comparison between both sets of data, as shown in **Figure 6a** in black and blue, respectively.

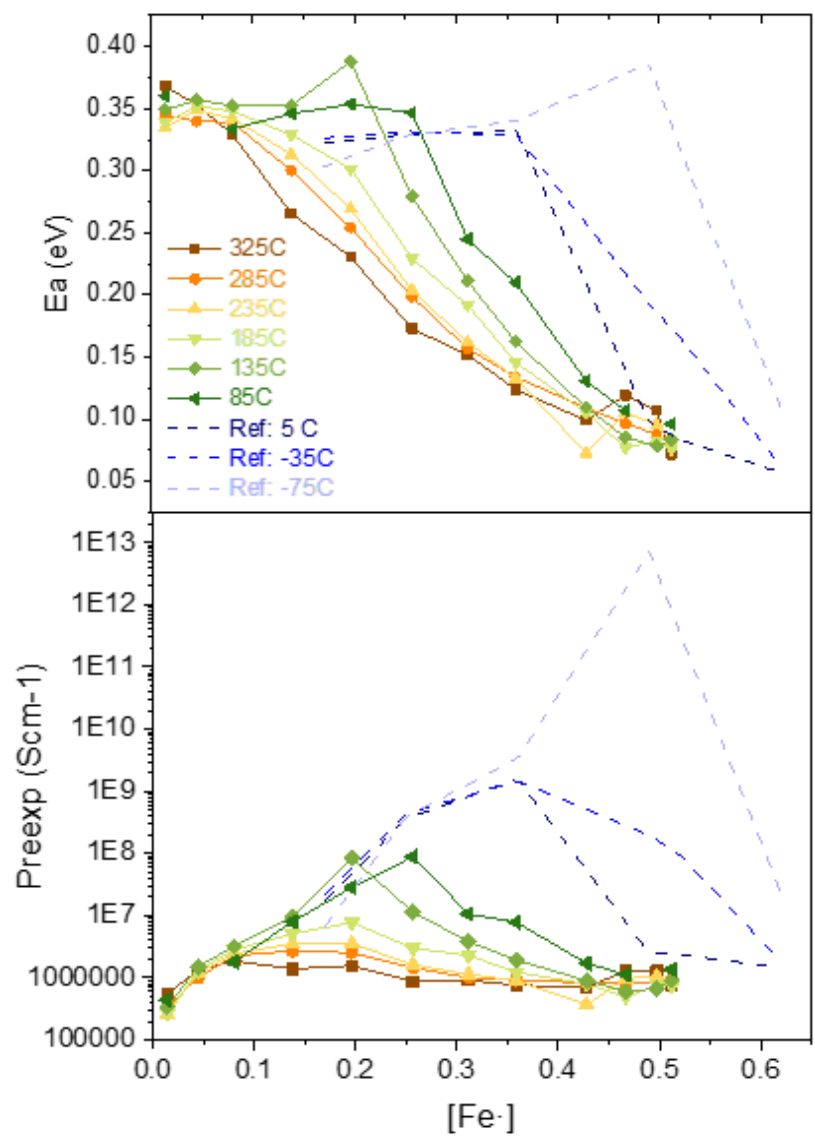

**Figure S11.** Comparison of the $E_a$ (up) and preexponential factor (down) from this paper and from Xie et al.[13] as a function of the $[Fe^{4+}]$

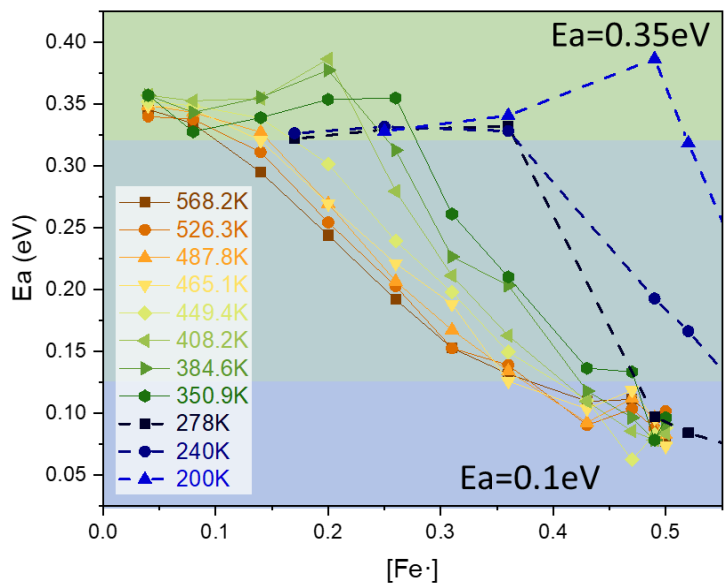

**Figure S12.** Figure ilustrating the $T_{Ea}$ calculation criteria showing the high activation area in green and the "lowered" $E_a$ area in blue

**SI7: General analysis on the properties of PV-LSF50**

**Figure S14** depicts a schematic of the Valence and Conduction Bands, VB and CB, respectively, of this family of materials in the PV phase. From left to right, the figure shows the effect of increasing the number of $Fe^{4+}$ holes in the lattice, aiming to represent the states with $[Fe^{4+}]=0$, $[Fe^{4+}]=0.5$ and $[Fe^{4+}]=1$, respectively. This $[Fe^{4+}]$ can be regulated via cationic doping in the A site or inducing oxygen vacancies ($V_O^{\bullet\bullet}$) changing the oxidation state as in this paper. As a simplification, the schematics in the figure can be interpreted as LSF50 in a stochiometric (center) and reduced state (left) but also $SrFeO_3$ (SFO) in its stochiometric (right), reduced (left) and half oxidized (center) states. It is known that in these materials the top of the VB consists of hybridized O2p and Fe3d eg orbitals, while unoccupied Fe 3d $t_{2g}$ and $e_g$ orbitals are thought to form the bottom of conduction band (CB).[9,11,16,17] As the material gets more oxidized, $Fe^{4+}$ holes are generated as part of the charge compensation process, causing the top of the VB to become partially unoccupied. This results in the emergence of an unoccupied Fe3d eg and O2p hybridized band, an acceptor state within the band gap that lies between the VB and the initial CB. Furthermore, with the rising concentration of holes, the degree of hybridization between the Fe3d and O2p orbitals becomes stronger, leading to a gradual reduction in the band gap. If the concentration of Fe4+ increases even more, the hole induced states can even cross the Fermi level, meaning that the material will act as a pure conductor. This can be seen as purely delocalized Fe4+, i.e., no polaron hopping could be applied to describe the conductivity anymore. This is actually the case for stochiometric SFO.

**Figure S14** also shows the variation of the optical transitions noted as A, B and C happening in the material. Transition A is correlated with the new possible transition from the edge of the VB to the hole-induced band. Remarkably, it can serve as an indicator for measuring the concentration of holes, which increases with the material oxidation.[8] Transitions B and C are related to the electron excitation from the hybridized Fe 3d/O2p orbitals in the VB toward the minority spin unoccupied $t_{2g}$ and $e_g$ orbitals in the CB, respectively.

A schematic of the interactions between iron sites of these three representative cases are also shown above the corresponding band diagrams in **Figure S14**. In the scenario where the concentration of Fe4+ holes is zero ([Fe3+] =1), the prevailing mechanism between Fe sites is Super-Exchange (SE). This mechanism induces Antiferromagnetic (AFM) spin ordering and hinders the electron jump to the adjacent site. This hindrance occurs because the probability of finding a neighboring hole is minimal, and the spin of the adjacent site is opposite, rendering the transition forbidden. With an increase in $[Fe^{4+}]$, this SE mechanism competes with Double

Exchange (DE) involving Fe3+ and adjacent Fe4+. As explained in the main text, this competition leads to a decrease in the Neel temperature ($T_N$).[18,19] If the spin ordering becomes non-AFM, either by surpassing $T_N$ or increasing the number of holes, the electron transition to the adjacent Fe site becomes more favorable. This results in a decrease in the activation energy ($E_a$) since the jump to the second-nearest-neighbor (SNN) will no longer be required for electron motion. If the hole concentration continues to increase, the valence band and the hole-induced level in the conduction band will approach each other, eventually closing the band gap. This transition turns the material into a metallic conductor, allowing electrons to jump almost freely from one $Fe^{4+}$ site to another, given the high probability that the orbital of the neighboring Fe is neither fully occupied nor has an opposite spin.

**Figure S15** illustrates the origin of optical and electronic transitions that can occur between Fe sites within the material. The left panel depicts the classic hopping example, where the electron moves to the nearest neighboring Fe with an associated activation energy ($E_a$). Similarly, the energy of optical transition A arises from the promotion of the electron to the level of the adjacent site, following the relationship explained by Holstein.[20]

However, when spins are aligned antiferromagnetically (AFM order in the Fe sublattice), the scenario is described by the right panel of **Figure S15**. Now, the jump to the adjacent Fe site is not favored, and the electron must transition to the second-nearest neighbor (SNN), requiring a higher $E_a$.[21] This also affects the optical transition since the promotion of the electron like in the left case is not allowed. In this scenario, the energy of the optical transition is higher because it must reach the energy of the SNN orbital. This is why both $E_a$ and the energy of transition A will decrease when the AFM order is disrupted. This disruption can occur either by raising the temperature (**Figure S7**) or by increasing $[Fe^{4+}]$ (**Figure S11 and S12**).

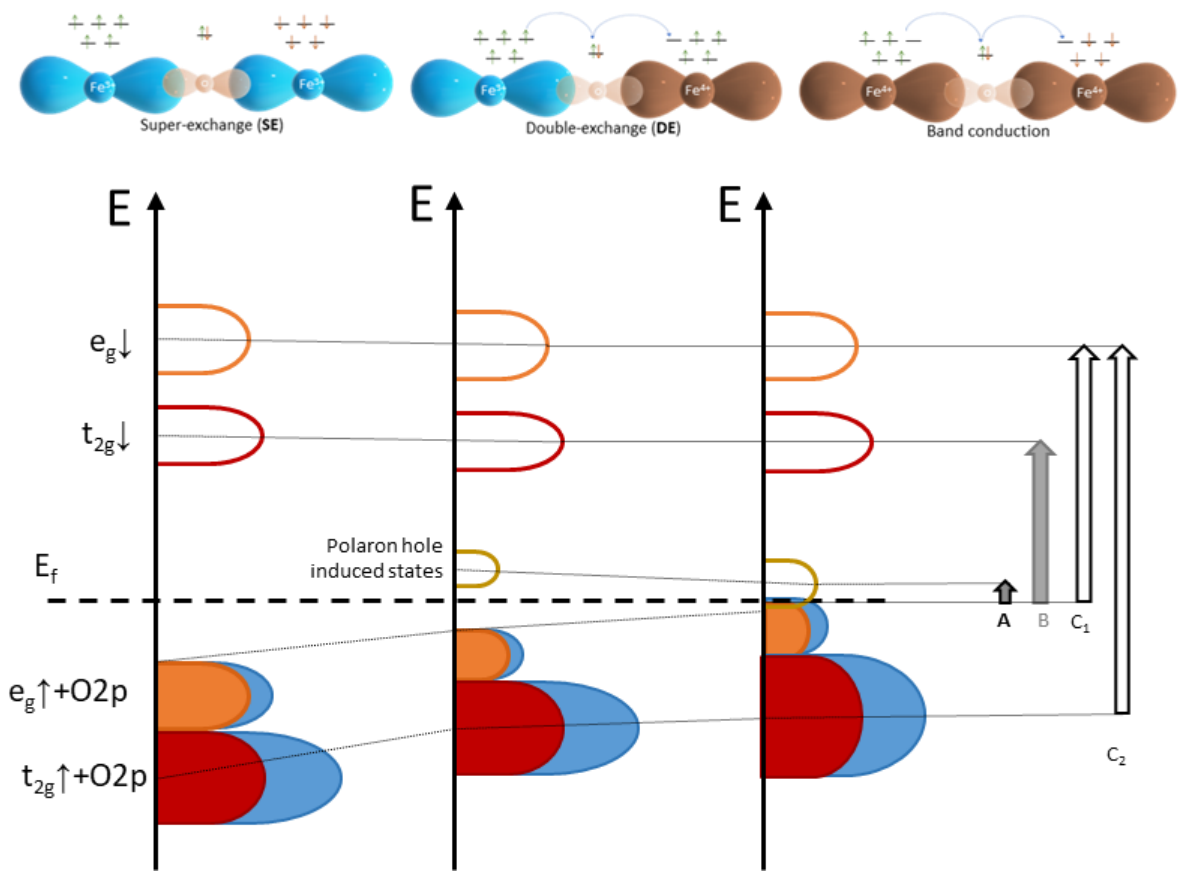

**Figure S13.** Schematic of the band change with the corresponding measured optical transitions and conduction mechanism for different $[Fe^{4+}]$ (=0, 0.5, 1; from left to right, respectively).

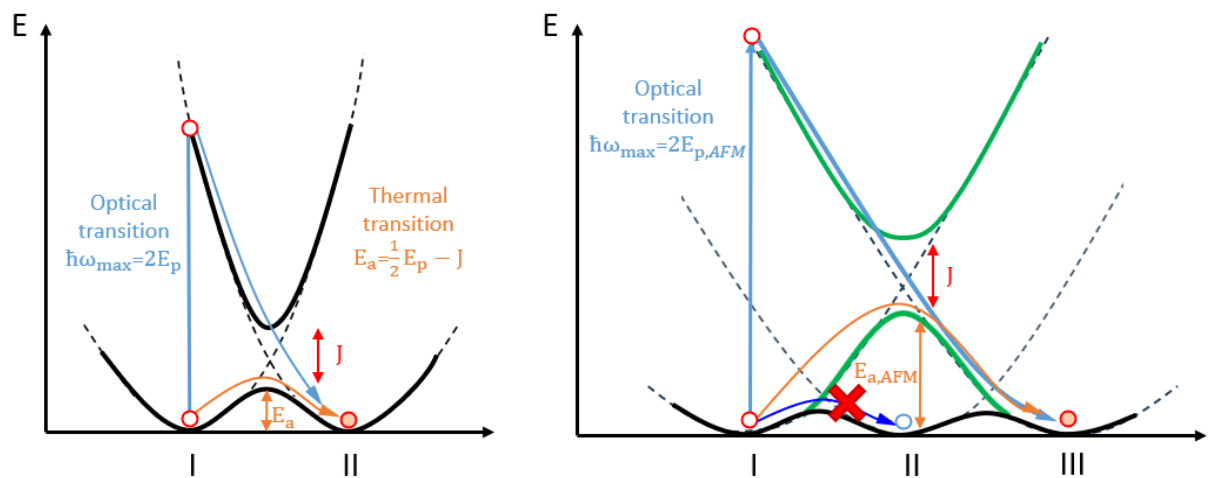

**Figure S14.** Figure illustrating the origin of the change in $E_{a,opt}$. Left figure shows the classical interpretation of optical transition of a polaron to the adjacent site. Right figure shows the case of an AFM ordering, where the hopping to the neighboring site is not allowed.

**SI8: Attempt of fitting LSF50 conductivity with the Bruggeman rule of mixing**

To completely discard phase formation or the mixing of phases as the origin of the exponential increase in conductivity for the low temperature regime we attempt to fit the central region conductivity with the Bruggeman's rule of mixing equation for a symmetric medium,[22] considering the sample as a local mixture of insulating and conductive volumes. To do so, the conductivity was divided in the three main regimes: low conducive phase, mixture of phases and, high conductive phase. The conductivity of each of the two extreme regimes was described using the adiabatic polaron hopping model as follows:

$$\sigma_{low}([Fe^{4+}],T) = A \cdot [Fe^{4+}] \cdot (1-[Fe^{4+}]) \cdot e^{-Ea_{low}/k_BT} \quad for\ [Fe^{4+}] < [Fe^{4+}]_{low} \tag{S10}$$

$$\sigma_{high}([Fe^{4+}],T) = B \cdot [Fe^{4+}] \cdot (1-[Fe^{4+}]) \cdot e^{-Ea_{high}/k_BT} \quad for\ [Fe^{4+}] > [Fe^{4+}]_{high} \tag{S11}$$

Where $\sigma_{low}$ and $\sigma_{high}$ are the conductivity in the low-conductive and the high-conductive stabilized regimes, respectively. Both conductivities depend on the hole concentration $[Fe^{4+}]$ and temperature $T$. A and B are proportionality factors, $k_B$ is the Boltzmann constant and, $Ea_{low} = 0.35eV$ and $Ea_{high} = 0.1eV$ are the activation energy of each of the stabilized phases. $[Fe^{4+}]_{BW}$ and $[Fe^{4+}]_{PV}$ are the composition of the BM and PV phases in the mixing, respectively. For the central region, Bruggeman's rule of mixing equation for a symmetric medium was used to obtain the conductivity, $\sigma_{mix}$, as a function of $[Fe^{4+}]$ and T:

$$f_{low} \cdot \frac{\sigma_{low}([Fe^{4+}]_{BW})-\sigma_{mix}}{\sigma_{low}([Fe^{4+}]_{BW})+\left(\frac{1-L}{L}\right)\cdot\sigma_{mix}} + f_{high} \cdot \frac{\sigma_{high}([Fe^{4+}]_{PV})-\sigma_{mix}}{\sigma_{high}([Fe^{4+}]_{PV})+\left(\frac{1-L}{L}\right)\cdot\sigma_{mix}} = 0 \tag{S12}$$

Where $\sigma_{low}([Fe^{4+}]_{low})$, $\sigma_{high}([Fe^{4+}]_{high})$ correspond to the conductivity of the low-conductive phase and the high-conductive phase dependent on their corresponding $[Fe^{4+}]$. The $L$ term is the depolarization coefficient. In this study, L was set at 0.5, that does also correspond to the case with percolation thresholds of 0.5. The volume fractions of each phase are represented by $f_{low}$ and $f_{high}$ and, they were calculated as:

$$f_{low} = \frac{[Fe^{4+}]_{high}-[Fe^{4+}]}{[Fe^{4+}]_{high}-[Fe^{4+}]_{low}} \tag{S13}$$

$$f_{high} = 1 - f_{low} = \frac{[Fe^{4+}]-[Fe^{4+}]_{low}}{[Fe^{4+}]_{high}-[Fe^{4+}]_{low}} \tag{S14}$$

**Figure S15** shows the fitting for T=50ºC on the LSF50. It can be clearly seen that the LSF50 fitting gave not good description of the data. These results also prove that LSF50 change in $E_a$ cannot be attributed to a structural phase transition or a mixture of two phases.

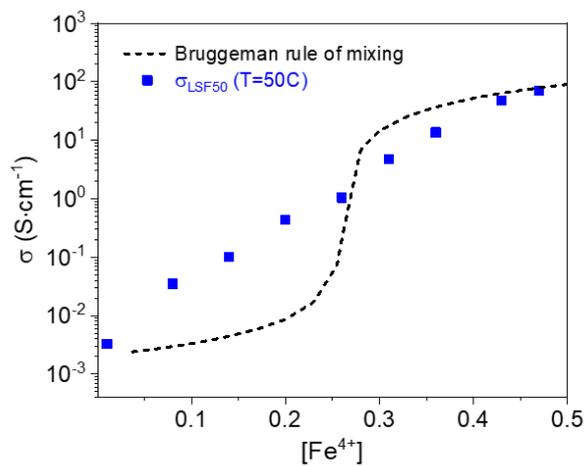

**Figure S15.** Bruggeman rule of mixing equation results for the conductivity of LSF50 at T=50ºC using the model presented in Equation S12

**SI9: Magnetic moment vs Temperature for LSF50 at different oxidation states**

Magnetic properties in these Transition Metal Oxides (TMOs) have a close relationship with electronic properties.[23,24] For this reason, different oxidation states ($[Fe^{4+}]$ =0.1, 0.2, 0.3, 0.45) were chosen to be studied under applied magnetic fields. The temperature dependence of the magnetization (M) was measured at a constant field (H=1kOe) and it is shown in **Figure S16a**. A small measurable magnetic signal could be observed. Nevertheless, as temperature was increased, a sudden change in the magnetization behavior was observed. This change in the M vs T slope was assigned to the antiferromagnetic-paramagnetic (AFM-PM) transition when overcoming the Néel Temperature ($T_N$). **Figure S16b** shows the $T_N$ of the sample as a function of $[Fe^{4+}]$ determined from the point at which the slope changes. The resulting values for the different oxidation states are included in **Figure 6a of the main text** where the change in slope happens. It can be inferred that the $T_N$ shifts to lower temperatures when the $[Fe^{4+}]$ increases. As expected for systems with competing Double-exchange (DE) and Superexchange (SE) interactions at all lattice bonds, spin fluctuations start to be enhanced.[18,19] In the case where SE dominates, an AFM state is expected with a $T_N$ dependent on the nearest-neighbor (NN) antiferromagnetic interaction $J_{AF} < 0$. This is the case of their parent compound, LFO (LaFeO$_3$) which has a $T_N \approx 740°C$. As the DE fraction increases, spins will not rigidly be oriented according to the predominantly AFM order. A simple way one can imagine this effect is thinking that an increase in increase of $[Fe^{4+}]$ diminishes SE interaction strength. Consequently, the effective AFM interaction strength, $J_{AF}^{eff}$, tends to 0, thus making the $T_N$ decrease, as reported by Wattiaux et al.[24]

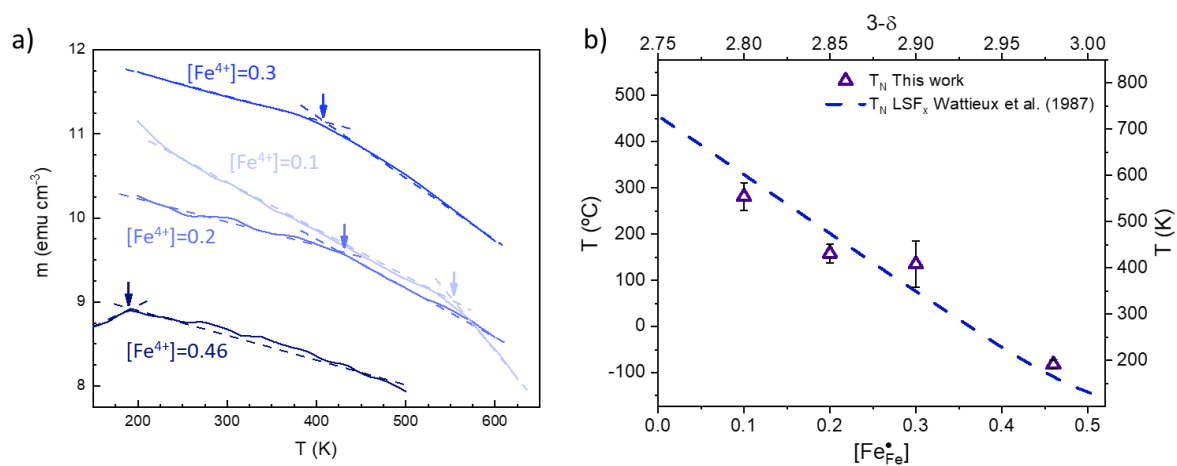

**Figure S16** a) Magnetization (M) vs. temperature (T) dependance measured at a constant 1kOe applied field for different oxidation states b) T$_N$ for the different oxidation states measured along with the results presented by Wattiaux et al.[24] for different LSFx as a function of their [$Fe^{4+}$].

**SI10: Co oxidation in the as-prepared sample**

After the in-plane voltage application and multi-state setting in the sample presented in **Section 2.6** in the main text, a 2.5-nm Co layer and 2 nm Pt layer were sputtered on top. In order to dismiss the Co oxidation and the possible $CoO_x$ influence in the Co loops, Co layer X-ray absorption spectroscopy (XAS) spectra of Co were measured in the as-deposited sample. **Figure S17** shows the L3 and L2 edge peaks of Co centered at approximately 777.6eV and 793eV, respectively. No evidence of $CoO_x$ presence was visible, discarding Co oxidation during the deposition process.

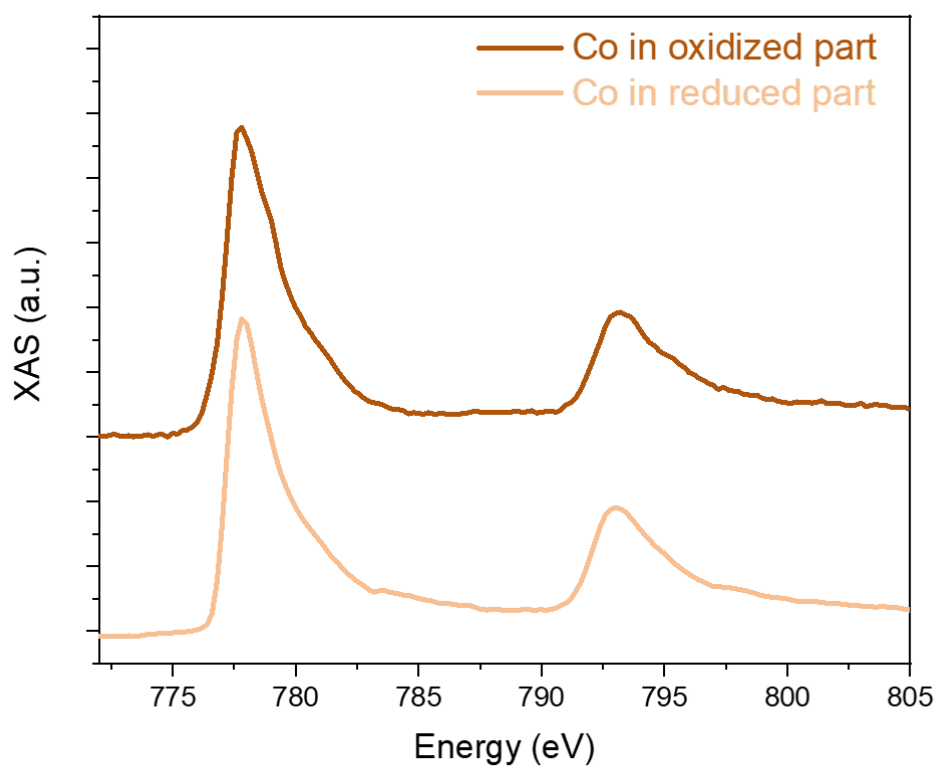

**Figure S17** XAS signal of Co deposited on top of the oxidized and reduced LSF50 parts of the sample